\documentclass[9pt,lineno]{style}

\usepackage{lipsum} 
\usepackage[version=4]{mhchem}
\usepackage{siunitx}
\usepackage{bm}
\usepackage{hyperref}
\usepackage{float}
\usepackage{amsmath,mathtools}
\newcommand{\bracket}[3]{\left#1 #3 \right#2}

\renewcommand{\b}{\bracket{(}{)}}

\newcommand{\Lbi}{\mathcal{L}_\text{BI}}

\newcommand{\Cv}{\mathcal{C}_\text{VI}}
\newcommand{\Cb}{\mathcal{C}_\text{BI}}

\newcommand{\Cvi}{\mathcal{C}_\text{VI;i}}
\newcommand{\Cbi}{\mathcal{C}_\text{BI;i}}

\DeclareSIUnit\Molar{M}
\DeclareMathOperator{\E}{E}

\let\P\relax
\DeclareMathOperator{\P}{P}
\DeclareMathOperator{\Q}{Q}
\DeclareMathOperator{\const}{const}

\DeclareUnicodeCharacter{2212}{-}
\DeclareMathOperator*{\tr}{tr}

\newcommand{\remove}[1]{}
\newcommand{\add}[1]{#1}
\newcommand{\replace}[2]{\remove{#1} \add{#2}}

\title{Signatures of Bayesian inference emerge from energy efficient synapses}

\usepackage{parskip}

\author[1*]{James Malkin}
\author[1,2]{Cian O'Donnell}
\author[1]{Conor Houghton}
\author[1]{Laurence Aitchison}
\affil[1]{Faculty of Engineering, University of Bristol, Bristol, UK}
\affil[2]{Intelligent Systems Research Centre, School of Computing, Engineering, and Intelligent Systems, Ulster University, Derry/Londonderry, UK}

\corr{james.malkin@bristol.ac.uk}{JM}

\begin{document}

\maketitle

\begin{abstract}

Biological synaptic transmission is unreliable, and this unreliability likely degrades neural circuit performance. 
While there are biophysical mechanisms that can increase reliability, for instance by increasing vesicle release probability, these mechanisms cost energy. 
We examined four such mechanisms along with the associated scaling of the energetic costs. 
We then embedded these energetic costs for reliability in artificial neural networks (ANN) with trainable stochastic synapses, and trained these networks on standard image classification tasks. 
The resulting networks revealed a tradeoff between circuit performance and the energetic cost of synaptic reliability. 
Additionally, the optimised networks exhibited two testable predictions consistent with pre-existing experimental data. 
Specifically, synapses with lower variability tended to have 1) higher input firing rates and 2) lower learning rates. 
Surprisingly, these predictions also arise when synapse statistics are inferred through Bayesian inference. 
Indeed, we were able to find a formal, theoretical link between the performance-reliability cost tradeoff and Bayesian inference. 
This connection suggests two incompatible possibilities: evolution may have chanced upon a scheme for implementing Bayesian inference by optimising energy efficiency, or alternatively, energy efficient synapses may display signatures of Bayesian inference without actually using Bayes to reason about uncertainty.

\end{abstract}

\section{Introduction}
\label{sec:intro}
The synapse is the major site of inter-cellular communication in the brain. 
The amplitude of synaptic post-synaptic potentials (PSPs) are usually highly variable or stochastic. This variability arises primarily presynaptically: the release of neurotransmitter from presynaptically-housed vesicles into the synaptic cleft has variable release probabilities and variable quantal sizes \citep{lisman1993quantal,branco2009probability,brock2020practical}.
Unreliable synaptic transmission seems puzzling, especially in light of evidence for low-noise, almost failure-free transmission at some synapses  \citep{paulsen1994quantal,paulsen1996quantal,bellingham1998developmental}.
Moreover, the degree to which a synapse is unreliable does not just vary from one synapse type to another, there is also an heterogeneity of precision amongst synapses of the same type \citep{murthy1997heterogeneous,dobrunz1997heterogeneity}. Given that there is capacity for more precise transmission, why is this capacity not used in more synapses?

Unreliable transmission degrades accuracy but \citet{laughlin1998metabolic} showed that the synaptic connection from a photoreceptor to a retinal large monopolar cell could increase its precision by increasing the number of synapses, averaging the noise away, but this comes at the cost of extra energy per bit of information transmitted. Moreover, \citet{levy2002energy} demonstrated that there is a value for the precision which optimises the energy cost of information transmission. In this paper, we explore this notion of a performance-energy tradeoff.

However, it is important to consider precision and energy cost in the context of neuronal computation; the brain does not simply transfer information from neuron to neuron, it performs computation through the interaction between neurons. However, models outlining a synaptic energy-performance tradeoff, \citep{laughlin1998metabolic,levy2002energy,goldman2004enhancement, harris2012synaptic,harris2019energy, karbowski2019metabolic}, predominantly consider information transmission between just two neurons and the corresponding information-theoretic view treats the synapse as an isolated conduit of information \citep{shannon1948mathematical}.
In contrast, in reality, a single synapse is just one unit of the computational machinery of the brain. 
As such, the performance of an individual synapse needs to be considered in the context of circuit performance. 
To perform computation in an energy-efficient way the circuit as a whole needs to allocate resources across different synapses to optimise the overall energy cost of computation \citep{yu2016energy,schug2021presynaptic}.

Here, we consider the consequences of a tradeoff between network performance and energetic reliability costs that depend explicitly upon synapse precision. We estimate the energy costs associated with precision by considering the biological mechanisms underpinning synaptic transmission. By including these costs in a neural network designed to perform a classification task, we observe a heterogeneity in synaptic precision and find that this ``allocation'' of precision is related to signatures of synapse ``importance'', which can be understood formally on the grounds of Bayesian inference.

\section{Results}

We proposed energetic costs for reliable synaptic transmission and then measured their consequences in an artificial neural network.

\subsection{Biophysical costs}

\begin{figure}[p]
    \centering
    \includegraphics[width=0.9\textwidth]{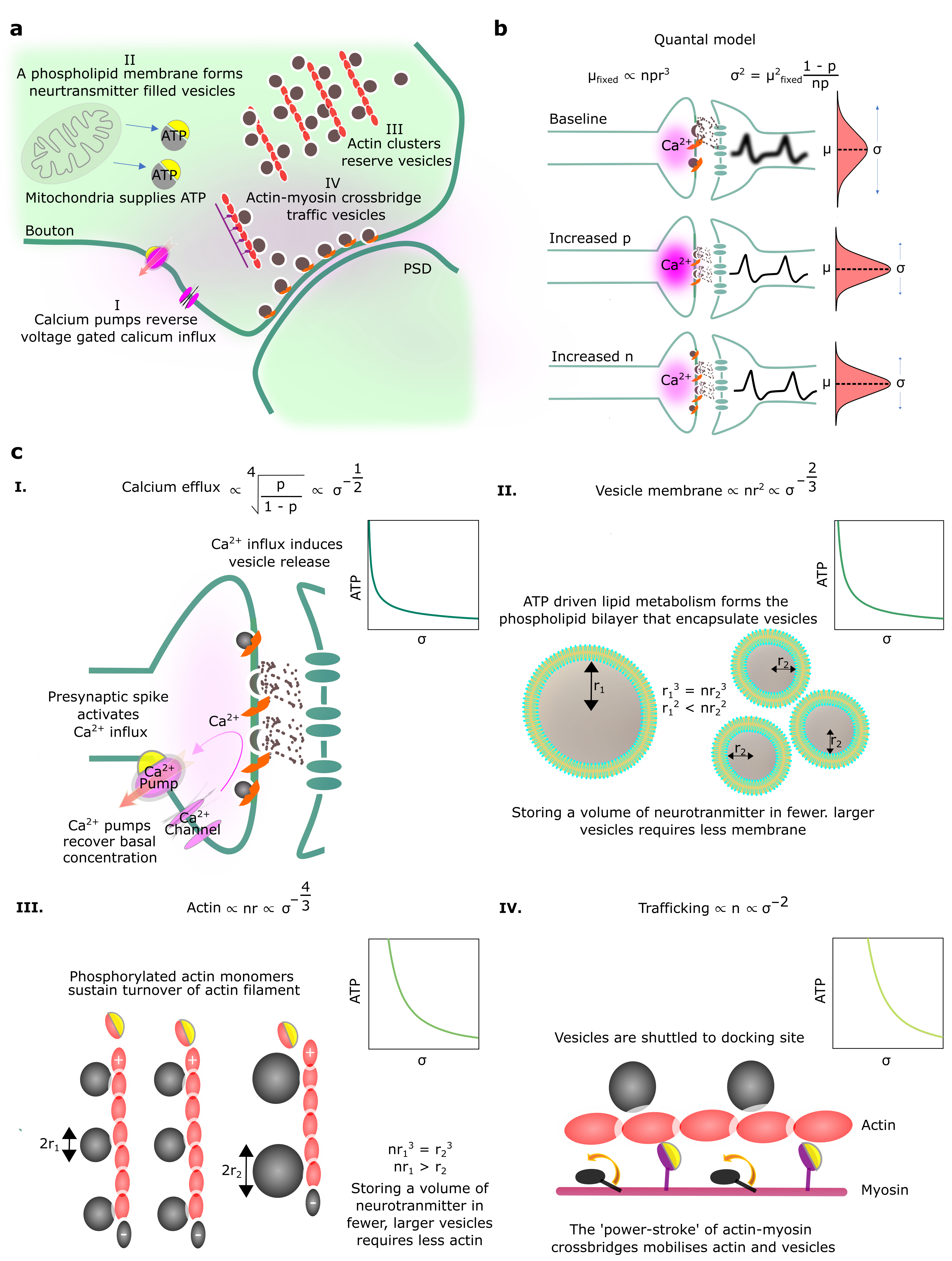}
\caption{\textbf{Physiological reliability costs} - \textbf{a.} Physiological processes that influence PSP precision. \textbf{b.} A binomial model of vesicle release. For fixed PSP mean, increasing $p$ or $n$ decreases PSP variance. We have substituted $q\propto r^3$ to reflect that vesicle volume scales quantal size \citep{karunanithi2002quantal}. \textbf{c.} Four different biophysical costs of reliable synaptic transmission. 
\textbf{I)} Calcium pumps reverse the calcium influx that triggers vesicle release. A high probability of vesicle release requires a large influx of calcium, and extruding this calcium is costly. \add{Note that $r$ represents the vesicle radius}.
\textbf{II)} 
An equivalent volume of neurotransmitter can be stored in few large vesicles or shared between many smaller vesicles. 
Sharing a fixed volume of neurotransmitter among many small vesicles reduces PSP variability but increases vesicle surface area, creating greater demand for phospholipid metabolism and hence greater energetic costs.   
\textbf{III)} Actin filament support the structure of vesicle clusters at the terminal. 
Many and large vesicles require more actin and higher rates of ATP dependent actin turnover.  
\textbf{IV)} There are biophysical costs that scale with the number of vesicles \citep{laughlin1998metabolic, attwell2001energy} e.g. vesicle trafficking driven by myosin-V active transport along actin filaments.
\label{fig1}
}
\end{figure}

Here, we seek to understand the biophysical energetic costs of synaptic transmission, and how those costs relate to the reliability of transmission (Fig.~\ref{fig1}a).
We start by considering the underlying mechanisms of synaptic transmission.
In particular, synaptic transmission begins with the arrival of a spike at the axon terminal.
This triggers a large influx of calcium ions into the axon terminal.
The increase in calcium concentration causes the release of neurotransmitter-filled vesicles docked at axonal release sites.
The neurotransmitter diffuses across the synaptic cleft to the postsynaptic dendritic membrane. 
There, the neurotransmitter binds with ligand-gated ion channels causing a change in voltage, i.e.\ a postsynaptic potential.
This process is often quantified using the \citet{katz1965measurement} quantal model of neurotransmitter release.
Under this model, for each connection between two cells, there are $n$ docked, readily releasable vesicles \add{(see Fig.~\ref{fig1}a for an illustration of a single synaptic connection with multi-vesicular release)}. 
\add{An alternative interpretation of this model might consider $n$ the number of uni-vesicular connections between two neurons.}
When the presynaptic cell spikes, each docked vesicle releases with probability $p$ and each released vesicle causes a postsynaptic potential of size $q$.
Thus, the mean, $\mu$, and variance, $\sigma^2$, of the PSP can be written (see Fig.~\ref{fig1}b),
\begin{align}
\mu &= n p q\cr
\sigma^2 &= n p (1-p) q^2.
\label{eq:binomial}
\end{align}
\add{where $q$ is considered a scaling variable. An assertion in our model is that variability in PSP strength is the result of variable numbers of vesicle release, not variability in $q$; here, during any PSP, $q$ is assumed constant across vesicles. While there is some suggestion that intra- and inter-site variability in $q$ is a significant component of PSP variability (see \citet{silver2003estimation}) we ultimately expect quantal variability to be small relative
to the variability attributed to vesicular release. This is supported by the classic observation that PSP amplitude histograms have a multi-peak structure \citep{boyd1956end, holler2021structure}; and by more direct measurement and modelling of vesicle release \citep{forti1997loose, raghavachari2004properties}.} 

We considered four biophysical costs associated with improving the reliability of synaptic transmission, while keeping the mean fixed, and derived the associated scaling of the energetic cost with PSP variance.

\textbf{Calcium efflux.}
Reliability is higher when the probability of vesicle release, $p$, is higher.
As vesicle release is triggered by an increase in intracellular calcium, greater calcium concentration implies higher release probability.
However, increased calcium concentration implies higher energetic costs.
In particular, calcium that enters the synaptic bouton will subsequently need to be pumped out. 
We take the cost of pumping out calcium ions to be proportional to the calcium concentration, and take the relationship between release probability and calcium concentration to be governed by a Hill Equation, following \citet{sakaba2001quantitative}.
The resulting relationship between energetic costs and reliability is $\text{cost} \propto \sigma^{-1/2}$ (Fig.~\ref{fig1}c (I);  
 see \hyperref[first:app]{Appendix - Reliability costs} for further details).

\textbf{Vesicle membrane surface area.}
There may also be energetic costs associated with producing and maintaining a large amount of vesicle membrane.
\citet{purdon2002energy} argues that phospholipid metabolism may take a considerable proportion of the brain's energy budget.
Additionally, costs associated with membrane surface area may arise because of leakage of hydrogen ions across vesicles \citep{pulido2021synaptic}.
Importantly, a cost for vesicle surface area is implicitly a cost on reliability.
In particular, we could obtain highly reliable synaptic release by releasing many small vesicles, such that stochasticity in individual vesicle release events averages out.
However, the resulting many small vesicles have a far larger surface area than a single large vesicle, with the same mean PSP.
Thus, a cost on surface area implies a relationship between energetic costs and reliability; in particular $\text{cost} \propto \sigma^{-2/3}$ (Fig.~\ref{fig1}c (II); see \hyperref[first:app]{Appendix - Reliability costs} for further details). 


\textbf{Actin.}
Another cost for small but numerous vesicles arises from a demand for structural organisation of the vesicles pool by filaments such as actin \citep{cingolani2008actin, gentile2022control}. 
Critically, there are physical limits to the number of vesicles that can be attached to an actin filament of a given length.
In particular, if vesicles are smaller we can attach more vesicles to a given length of actin, but at the same time, the total vesicle volume (and hence the total quantity of neurotransmitter) will be smaller (Fig.~\ref{fig1}c (III)).
A fixed cost per unit length of actin thus implies a relationship between energetic costs and reliability of, $\text{cost}\propto \sigma^{-4/3}$ (see \hyperref[first:app]{Appendix - Reliability costs}).

\textbf{Trafficking.}
A final class of costs is proportional to the number of vesicles \citep{laughlin1998metabolic}.
One potential biophysical mechanism by which such a cost might emerge is from active transport of vesicles along actin filaments or microtubles to release sites \citep{chenouard2020synaptic}. 
In particular, vesicles are transported by ATP-dependent myosin-V motors \citep{bridgman1999myosin}, so more vesicles require a greater energetic cost for trafficking.
Any such cost proportional to the number of vesicles gives rise to a relationship between energetic cost and PSP variance of the form, $\text{cost} \propto \sigma^{-2}$ (Fig.~\ref{fig1}c (IV); see \hyperref[first:app]{Appendix - Reliability costs}).

\textbf{Costs related to PSP mean/magnitude}
While costs on precision are the central focus of this paper, it is certainly the case that other costs relating to the mean PSP magnitude constitute a major cost of synaptic transmission. 
For example, high amplitude PSPs require a large quantity of neurotransmitter, high probability of vesicle release, and a large number of post-synaptic receptors \citep{attwell2001energy}. 
These can be formalised as costs on the PSP mean, $\mu$, and can additionally be related to L1 weight decay in a machine learning context \citep{rosset2006sparse, sacramento2015energy}. 

%


\subsection{Reliability costs in artificial neural networks}
Next, we sought to understand how these biophysical energetic costs of reliability might give rise to patterns of variability in a trained neural network.
Specifically, we trained artificial neural networks (ANNs) using an objective that embodied a tradeoff between performance and reliability costs,
\begin{align}
\text{overall cost} = \text{performance cost} + \text{magnitude cost} + \text{reliability cost}.
\label{eq:wordobjective}
\end{align}
The ``performance cost" term measures the network's performance on the task, for instance in our classification tasks we used the usual cross-entropy cost.
The ``magnitude cost'' term captures costs that depend on the PSP mean, while the ``reliability cost'' term captures costs that depend on the PSP precision.
In particular,
\begin{align}
\label{eq:magnitudeCost}
\text{magnitude cost} &= \lambda\sum_i |\mu_i|,\\
\label{eq:reliabilityCost}
\text{reliability cost} &= c\sum_i\sigma_i^{-\rho}.
\end{align} 
Here, $i$ indexes synapses, and recall that $\sigma_i$ is the standard deviation of the $i$th synapse.
%
%
The multiplier $c$ in the reliability cost determines the strength of the reliability cost relative to the performance cost. Small values for $c$ imply that the reliability cost term is less important, permitting precise transmission and higher performance. Large values for $c$ give greater importance to the reliability cost encouraging energy efficiency by allowing higher levels of synaptic noise, causing detriment to performance (see Fig.~\ref{fig:accuracy}).

We trained fully-connected, rate-based neural network to classify MNIST digits.
Stochastic synaptic PSPs were sampled from a Normal distribution,
\begin{equation}
w_i \sim \text{Normal}(\mu_i, \sigma_i).
\end{equation}
where, recall, $\mu_i$ is the PSP mean and $\sigma_i^2$ is the PSP variance for the $i$th synapse.
The output firing rate was given by,
\begin{align}
\text{firing rate} = f\left( \sum_{i} w_i x_i - w_0 \right). 
\end{align}
Here, $\sum_{i} w_i x_i - w_0$ can be understood as the somatic membrane potential, and $f$ represents the relationship between somatic membrane potential and firing rate; we used ReLU \citep{fukushima1975cognitron}.
We optimised network parameters $\mu_i$ and $\sigma_i$ using Adam \citep{kingma2014adam} (see \hyperref[sec:methods]{Methods} for details on architecture and hyperparameters).

\subsection{The tradeoff between accuracy and reliability costs in trained networks}
Next we sought to understand how the tradeoff between accuracy and reliability cost manifests in trained networks.
Perhaps the critical parameter in the objective, (Eq.~\ref{eq:wordobjective} and Eq.~\ref{eq:reliabilityCost}) was $c$, which controlled the importance of the reliability cost relative to the performance cost.
We trained networks with a variety of different values of $c$, and with four values for $\rho$ motivated by the biophysical costs (the different columns). 

\add{In practice all the reliability costs and others we may have overlooked should together constitute an overall energetic reliability cost. However, it is difficult to estimate the specific contributions of different costs, that is the individual values of $c$. While \citet{attwell2001energy, engl2015non} estimate the ATP demands for various synaptic processes, it is difficult to relate these to the relative scale of each cost at a synapse level. Therefore, for simplicity, we kept each cost separate, training neural networks with just one choice of reliability cost; emphasising results shared across all costs. It is possible that one cost dominates all the others, but if that is not the case it will be necessary to use a more complicated reliability cost. However, since we have considered four costs with very different power-law behaviours, it is likely the behaviour will not be significantly different to what we have observed.}

\begin{figure}
    \centering
    \includegraphics{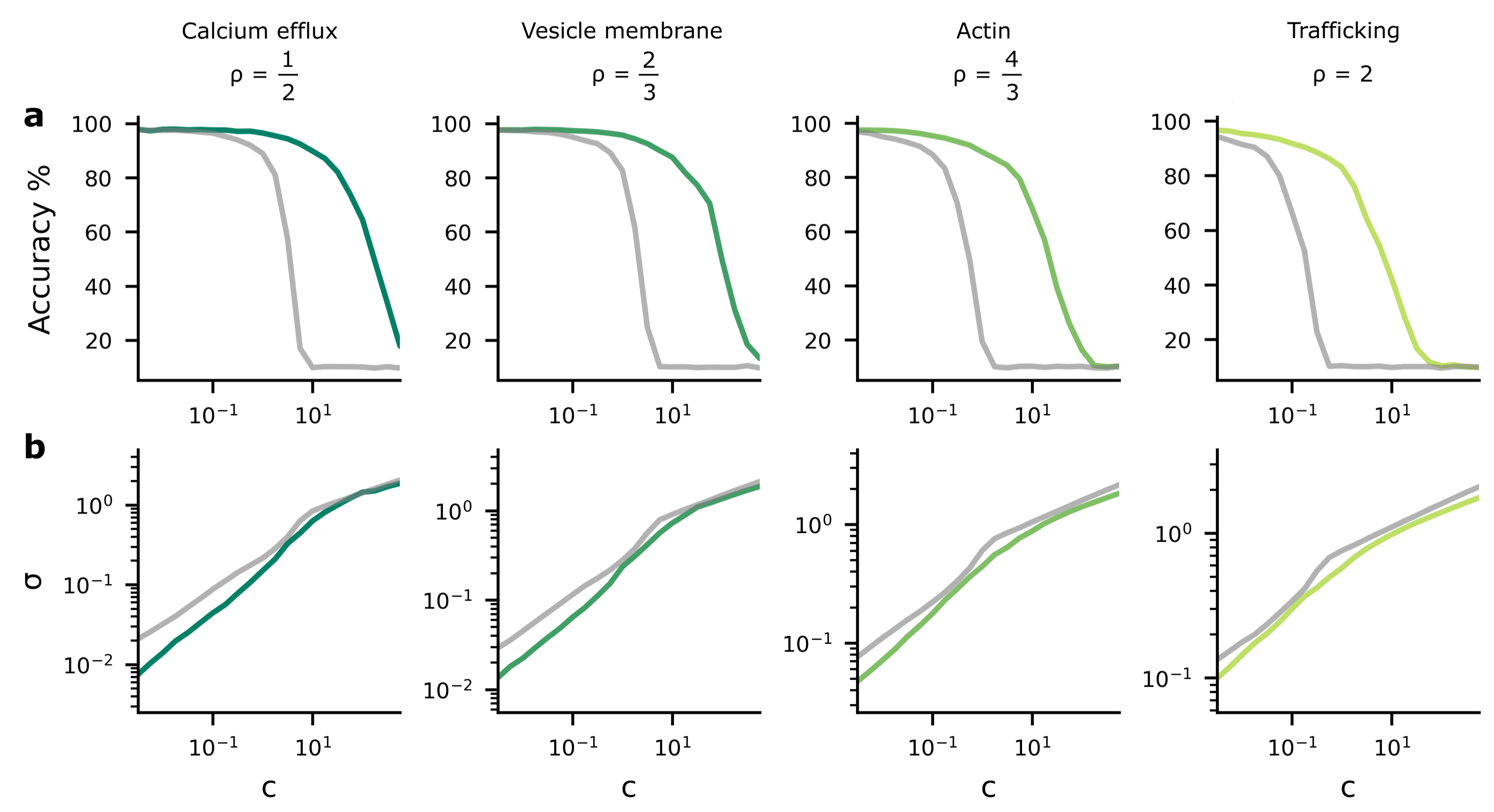}
    \caption{\textbf{Accuracy and PSP variance as we change the tradeoff between reliability and performance costs.} We changed the tradeoff by modifying $c$, in Eq.~\ref{eq:reliabilityCost}, which multiplies the reliability cost. \textbf{a.} As the reliability cost multiplier, $c$, increases, the accuracy decreases considerably. The green lines show the heterogeneous noise setting where the noise level is optimised on a per-synapse basis, while the grey lines show the homogeneous noise setting, where the noise is optimised, but forced to be the same for all synapses. \textbf{b.} When the reliability cost multiplier, $c$ increases, the synaptic noise level (specifically, the average standard deviation, $\sigma$) increases.
    }
    \label{fig:accuracy}

\end{figure}

As expected, we found that as $c$ increased, performance fell (Fig.~\ref{fig:accuracy}a) and the average synaptic standard deviation increased (Fig.~\ref{fig:accuracy}b).
Importantly, we considered two different settings. 
First, we considered an homogeneous noise setting, where $\sigma_i$ is optimised but kept the same across all synapses (grey lines).
Second, we considered an heterogeneous noise setting, where $\sigma_i$ is allowed to vary across synapses, and is optimised on a per-synapse basis.
We found that heterogeneous noise (i.e.\ allowing the noise to vary on a per-synapse basis) improved accuracy considerably for a fixed value of $c$, but only reduced the average noise slightly.

\begin{figure}[h!]
    \centering
    \includegraphics{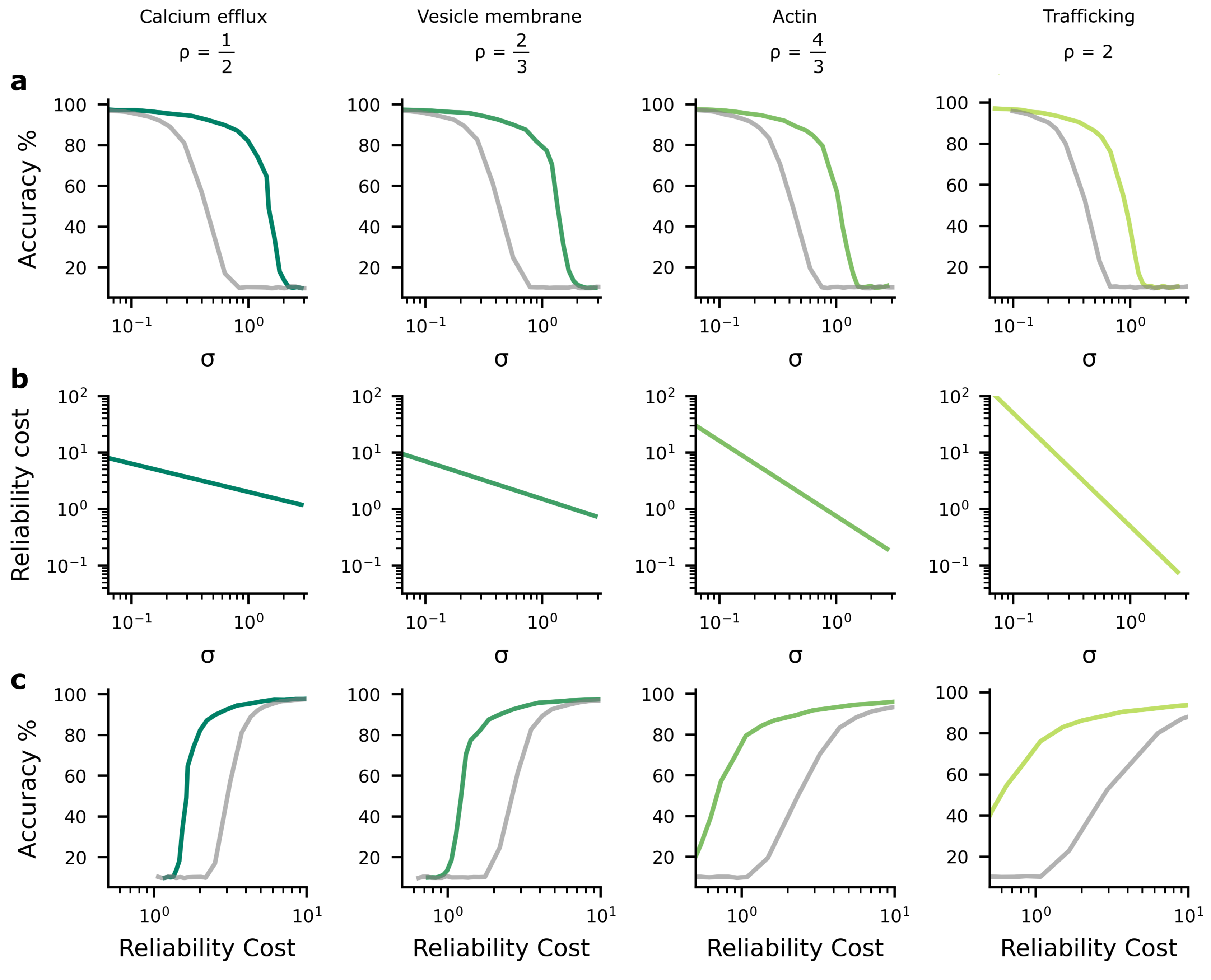}
    \caption{\textbf{The performance-reliability cost tradeoff in ANN simulations} - \textbf{a.} Accuracy decreases as the average PSP standard deviation, $\sigma$, increases. 
    The grey lines are for the homogeneous noise setting where the PSP variance is optimised but isotropic (i.e.\ the same across all synapses), while the green lines is for the heterogeneous noise setting, where the PSP variances are optimised individually on a per-synapse basis.
    \textbf{b.} Increasing reliability by reducing $\sigma^2$ leads to greater reliability costs, and this relationship is different for different biophysical mechanisms and hence values for $\rho$ (columns).
    \textbf{c.} Higher accuracy therefore implies larger reliability cost.}
    \label{fig:performance}
\end{figure}

The findings in Fig.~\ref{fig:accuracy} imply a tradeoff between accuracy and average noise level, $\sigma$, as we change $c$.
If we explicitly plot the accuracy against the noise level using the data from Fig.~\ref{fig:accuracy}, we see that as the synaptic noise level increases, the accuracy decreases (Fig.~\ref{fig:performance}a).
Further, the synaptic noise level is associated with a reliability cost (Fig.~\ref{fig:performance}b), and this relationship changes in the different columns as they use different values of $\rho$ associated with different biological mechanisms that might give rise to the dominant biophysical reliability cost.
Thus, there is also a relationship between accuracy and reliability costs (Fig.~\ref{fig:performance}c), with accuracy increasing as we allow the system to invest more energy in becoming more reliable, which implies a higher reliability cost.
Again, we plotted both the homogeneous (grey lines) and heterogeneous noise cases (green lines).
We found that heterogeneous noise allowed for considerably improved accuracy at a given average noise standard deviation or a given reliability cost.

\begin{figure}[h!]
    \centering
\includegraphics{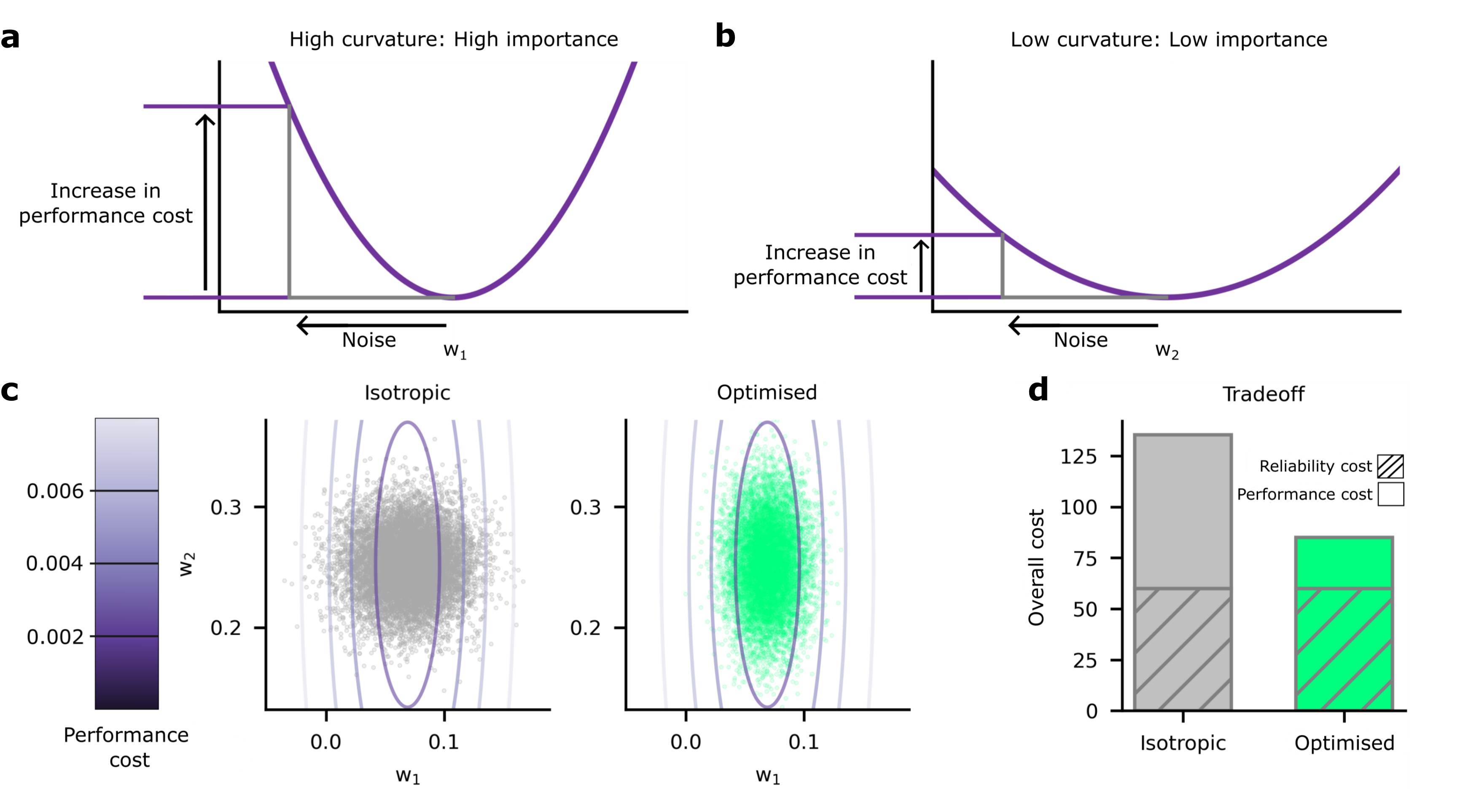}
    \caption{\textbf{Schematic depiction of the impact of synaptic noise on synapses with different importance.} \textbf{a.} First, we considered an important synapse for which small deviations in the weight, $w_1$, e.g.\ driven by noise, imply a large increase in the performance cost. This can be understood as a high curvature of the performance cost as a function of $w_1$. \textbf{b.} Next we considered an unimportant synapse, for which deviations in the weights cause far less increase in performance cost. \textbf{c.} A comparison of the impacts of homogeneous and optimised heterogeneous variability for synapses $w_1$ and $w_2$ from a and b. The performance cost is depicted using the purple contours, and realisations of the PSPs driven by synaptic variability are depicted in the grey/green points. The grey points (left) depict homogeneous noise while the green points (right) depict optimised, heterogeneous noise. \textbf{d.} The noise distributions in panel c are chosen to keep the same reliability cost (diagonally hatched area); but the homogeneous noise setting has far a higher performance cost, primarily driven by larger noise in the important synapse, $w_1$.
    \label{fig:curve}
}
\end{figure}

\begin{figure}[h!]
    \centering
    \includegraphics{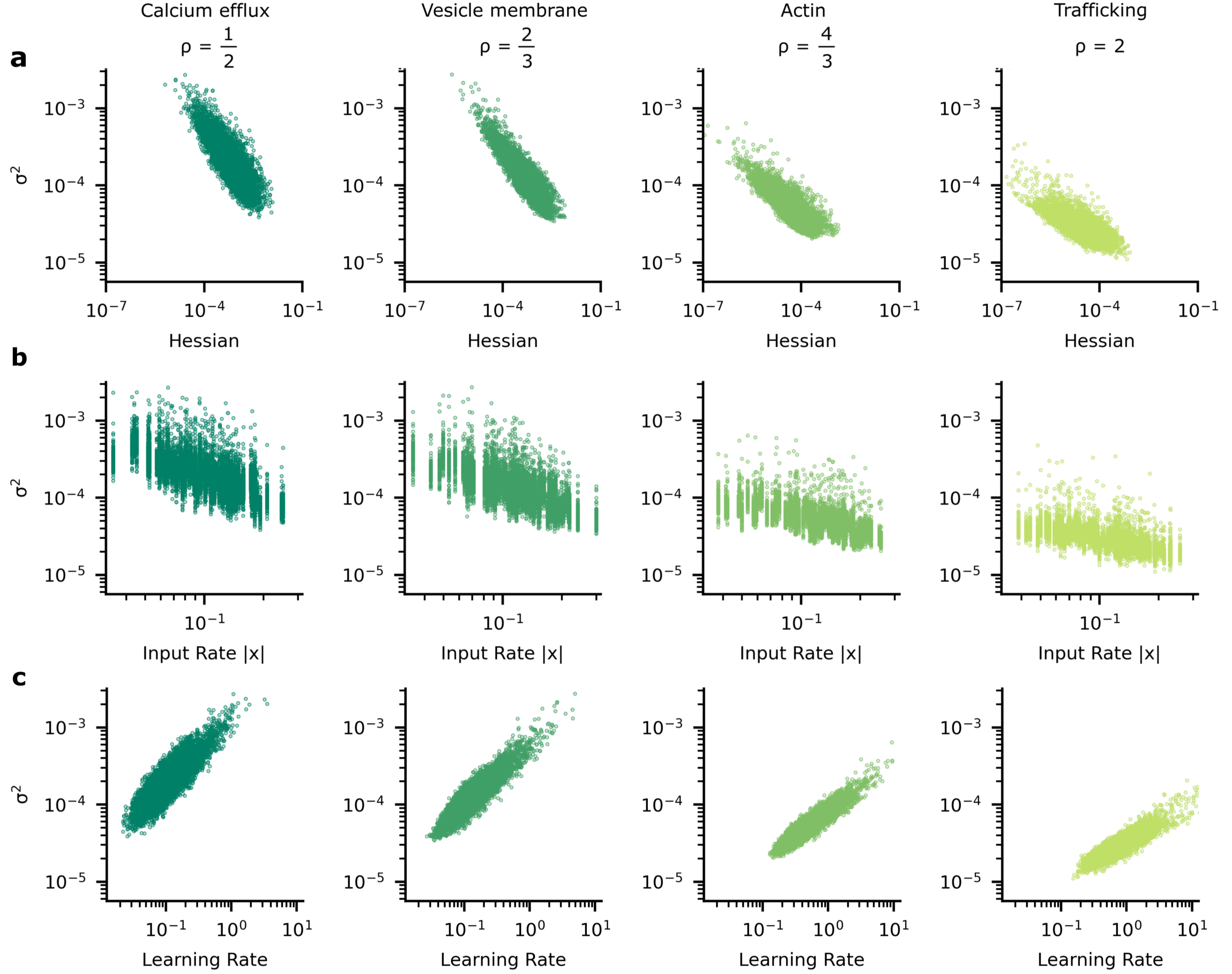}
    \caption{\textbf{The heterogeneous patterns of synapse variability in ANNs optimised by the tradeoff} - We present data patterns on logarithmic axis between signatures of synapse importance and variability for 10,000 (100 neuron units, each with 100 synapses) synapses that connect two hidden layers in our ANN.  \textbf{a.} Synapses whose corresponding diagonal entry in the Hessian is large have smaller variance.
    \textbf{b.} Synapses with high variance have faster learning rates 
    \textbf{c.} As input firing rate increases, synapse variance decreases.
    \label{fig:optimising}}
\end{figure}

\subsection{Energy-efficient patterns of synapse variability}
We found that the heterogeneous noise setting, where we individually optimise synaptic noise on a per-synapse basis, performed considerably better than the homogeneous noise setting (Fig.~\ref{fig:performance}).
This raised an important question: how does the network achieve such large improvements by optimising the noise levels on a per-synapse basis?
We hypothesised that the system invests a lot of energy in improving the reliability for ``important'' synapses, i.e.\ synapses whose weights have a large impact on predictions and accuracy (Fig.~\ref{fig:curve}a).
Conversely, the system allows unimportant synapses to have high variability, which reduces reliability costs (Fig.~\ref{fig:curve}b). 
To get further intuition, we compared both $w_1$ and $w_2$ on the same plot (Fig.~\ref{fig:curve}c).
Specifically, we put the important synapse, $w_1$ from Fig.~\ref{fig:curve}a, on the horizontal axis, and the unimportant synapse, $w_2$ from Fig.~\ref{fig:curve}b, on the vertical axis.
In Fig.~\ref{fig:curve}c, the relative importance of the synapse is now depicted by how the cost increases as we move away from the optimal value of the weight.
Specifically, the cost increases rapidly as we move away from the optimal value of $w_1$, but increases much more slowly as we move away from the optimal value of $w_2$.
Now, consider deviations in the synaptic weight driven by homogeneous synaptic variability (Fig.~\ref{fig:curve}c left, grey points).
Many of these points have poor performance (i.e.\ a high performance cost), due to relatively high noise on the important synapse (i.e.\ $w_1$).
Next, consider deviations in the synaptic weight driven by heterogeneous, optimised variability (Fig.~\ref{fig:curve}c left, green points).
Critically, optimising synaptic noise reduces variability for the important synapse, and that reduces the average performance cost by eliminating large deviations on the important synapse.
Thus, for the same overall reliability cost, heterogeneous, optimised variability can achieve much lower performance costs, and hence much lower overall costs than homogeneous variability (Fig.~\ref{fig:curve}d).

To investigate experimental predictions arising from optimised, heterogeneous variability, we needed a way to formally assess the ``importance'' of synapses.
We used the ``curvature'' of the performance cost: namely the degree to which small deviations in the weights from their optimal values will degrade performance.
If the curvature is large (Fig.~\ref{fig:curve}a), then small deviations in the weights, e.g.\ those caused by noise, can drastically reduce performance.
In contrast, if the curvature is smaller (Fig.~\ref{fig:curve}b), then small deviations in the weights cause a much smaller reduction in performance.
As a formal measure of the curvature of the objective, we used the Hessian matrix, $\bm{H}$. This describes the shape of the objective as a function of the synaptic weights, the $w_i$s: specifically, it is the matrix of second derivatives of the objective, with respect to the weights, and measures the local curvature of objective. We were interested in the diagonal elements, $H_{ii}$; the second derivatives of the objective with respect to $w_i$.


We began by looking at how the optimised synaptic noise varied with synapse importance, as measured by the curvature or, more formally, the Hessian (Fig.~\ref{fig:optimising}a). 
\add{The Hessian values were estimated using the average-squared gradient, see \hyperref[third:app]{Appendix - Synapse importance and gradient magnitudes}.} We found that as the importance of the synapse increased, the optimised noise level decreased.
These patterns of synapse variability make sense because noise is more detrimental at important synapses and so it is worth investing energy to reduce the noise in those synapses.

%
However, this relationship (Fig.~\ref{fig:optimising}a) between the importance of a synapse and the synaptic variability is not experimentally testable, as we are not able to directly measure synapse importance.
That said, we are able to obtain two testable predictions.
First, the input rate in our simulations was negatively correlated with optimised synaptic variability (Fig.~\ref{fig:optimising}b).
Second, the optimised synaptic variability was larger for synapses with larger learning rates (Fig.~\ref{fig:optimising}c).
\add{Critically, similar patterns have been observed in experimental data. 
In Fig.~\ref{fig:exp}a we present the negative correlation between learning rate and synaptic reliability presented by \citet{schug2021presynaptic} from in-vitro measurements of V1 (layer 5) pyramidal synapses before and after STDP induced LTP conducted by \citet{sjostrom2001rate}. 
Furthermore, a relationship between input firing rate and synaptic variability was observed by \citet{aitchison2021synaptic} using in-vivo functional recordings from V1 (layer 2/3) \citep{ko2013emergence} (Fig.~\ref{fig:exp}b).}

To understand why these patterns of variability emerge in our simulations and in data, we need to understand the connection between synapse importance, synaptic inputs (Fig.~\ref{fig:optimising}b, Fig.~\ref{fig:exp}b) and synaptic learning (Fig.~\ref{fig:optimising}c, Fig.~\ref{fig:exp}a).
Perhaps the easiest connection is between the synapse importance and the input firing rate.
If the input cell never fires, then the synaptic weight cannot affect the network output, and the synapse has zero importance (and also zero Hessian (see \hyperref[second:app]{Appendix - High input rates and high precision at important synapses})).
This would suggest a tendency for synapses with higher input firing rates to be more important, and hence to have lower variability.
This pattern is indeed borne out in our simulations (Fig.~\ref{fig:optimising}b; also see \hyperref[sixth:app]{Supplementary - Appendix 6--Fig.~\ref{fig:all_layers}}), though of course there is a considerable amount of noise: there are a few important synapses with low input rates, and vice-versa.


Next, we consider the connection between learning rate and synapse importance. 
To understand this connection, we need to choose a specific scheme for modulating the learning rate as a function of the inputs. 
\add{While the specific scheme for modulating the learning rate is ultimately an assumption, we believe modern deep learning offers strong guidance as to the optimal family of schemes for modulating the learning rate.}
In particular, modern, state-of-the-art, update rules for artificial neural networks almost always use an adaptive learning rate. 
These adaptive learning rates, $\eta_i$, (including the most common such as Adam and variants)  almost always use a normalising learning rate which decreases in response to high incoming gradients,
\begin{align}
  \label{eq:adaptive_learning_rate}
  \eta_i &= \frac{\eta_\text{base}}{\sqrt{\langle g_i^2 \rangle}}.
\end{align}
Specifically, the local learning rate for the $i$th synapse, $\eta_i$, is usually a base learning rate, $\eta_\text{base}$, divided by the root-mean-square gradient at this synapse $\sqrt{\langle g_i^2 \rangle}$. Critically, the root-mean-square gradient turns out to be strongly related to synapse importance. Intuitively, important synapses with greater impact on network predictions will have larger gradients (see \hyperref[third:app]{Appendix - Synapse importance and gradient magnitudes}). 

In-vivo performance requires selective formation, stabilisation and elimination of long term plasticity (LTP) \citep{yang2009stably}, raising the questions as to which biological mechanisms are able to provide this selectivity. Reducing updates at historically important synapses is one potential approach to determining which synapses should have their strengths adjusted and which should be stabilised. 
Adjusting learning rates based on synapse importance enables fast, stable learning \citep{lecun2002efficient, kingma2014adam, khan2018fast, aitchison2020bayesian,  martens2020new, jegminat2022learning}.  

For our purposes, the crucial point is that when training using an adaptive learning rate such as Eq.~\ref{eq:adaptive_learning_rate}, important synapses have higher root-mean-squared gradients, and hence lower learning rates.
Here we use a specific set of update rules which uses this adaptive learning rate (i.e.\ Adam \citep{kingma2014adam,yang2022synaptic}).
Thus, we can use learning rate as a proxy for importance, allowing us to obtain the predictions tested in Fig.~\ref{fig:optimising}b which match Fig.~\ref{fig:optimising}a/c. 

\subsection{The connection to Bayesian inference}
\begin{figure}
    \centering
    \includegraphics{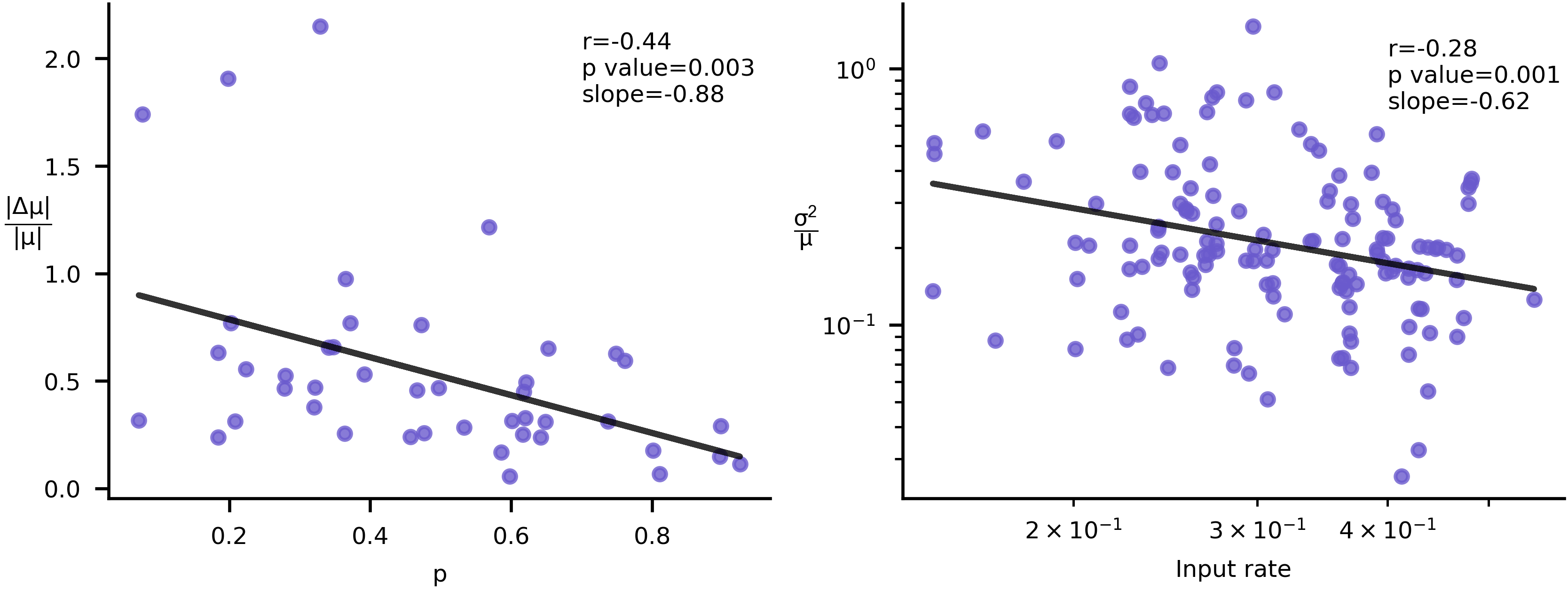}
    \caption{\textbf{Experimental signatures of Bayesian synapses} -  The Bayesian synapse hypothesis predicts relationships between synapse reliability, learning rate and input rate.  \textbf{a}) Synapses with higher probability of release, $p$ demonstrate smaller increases in synaptic mean following LTP induction. This pattern was originally observed by \citet{schug2021presynaptic}). 
    \textbf{b)} As input firing rates are increased, normalised EPSP variability decreases with $slope= -0.62$ \citep{aitchison2021synaptic}).
    \label{fig:exp}}
\end{figure}

Surprisingly, our experimental predictions obtained for optimised, heterogeneous synaptic variability (Fig.~\ref{fig:optimising}) match those arising from Bayesian synapses presented in Fig.~\ref{fig:exp} (i.e.\ synapses that use Bayes to infer their weights \citep{aitchison2021synaptic}). 
Our first prediction was that lower variability implies a lower learning rate.
The same prediction also arises if we consider Bayesian synapses.
In particular, if variability and hence uncertainty is low, then a Bayesian synapse is very certain that it is close to the optimal value.
In that case, new information should have less impact on the synaptic weight, and the learning rate should be lower.
Our second prediction was that higher presynaptic firing rates imply less variability.
Again, this arises in Bayesian synapses: Bayesian synapses should become more certain and less variable if the presynaptic cell fires more frequently. 
Every time the presynaptic cell fires, the synapse gets a feedback signal which gives a small amount of information about the right value for that synaptic weight.
So the more times the presynaptic cell fires, the more information the synapse receives, and the more certain it becomes.

This match between observations for our energy-efficient synapses and previous work on Bayesian synapses led us to investigate potential connections between energy efficiency and Bayesian inference.
Intuitively, there turns out to be a strong connection between synapse importance and uncertainty. 
Specifically, if a synapse is very important, then the performance cost changes dramatically when there are errors in that synaptic weight.
That synapse therefore receives large gradients, and hence strong information about the correct value, rapidly reducing uncertainty.

\begin{figure}[h!]
    \centering
    \includegraphics{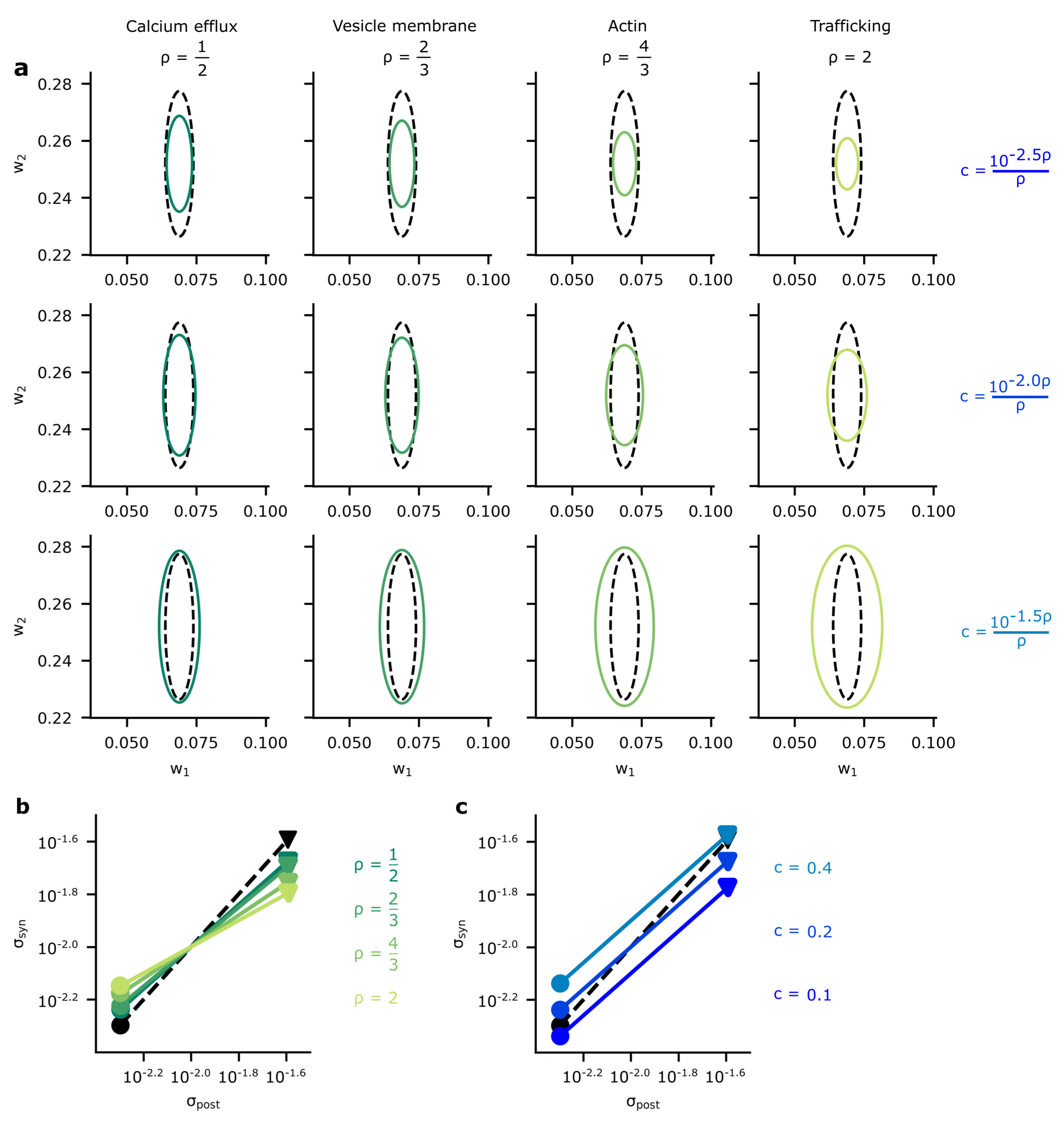}
    \caption{\textbf{A comparison of optimised synaptic variability and posterior variance.} - \textbf{a.} Posterior variance (grey-dashed ellipses) plotted alongside optimised synaptic variability (green ellipses) for different values of $\rho$ (columns) and $c$ (rows) \add{for an illustrative pair of synapses}. Note that using fixed values of $c$ for different $\rho$'s dramatically changed the scale of the ellipses. Instead, we chose $c$ as a function of $\rho$ to ensure that the scale of the optimised noise variance was roughly equal across different $\rho$.  This allowed us to highlight the key pattern: that smaller values for $\rho$ give optimised variance closer to the true posterior variances, while higher values for $\rho$ tended to make the optimised synaptic variability more isotropic. 
    \textbf{b.} To understand this pattern more formally, we plotted the synaptic variability as a function of the posterior variance for different values of $\rho$. Note that we set $c$ to $c=\tfrac{10^{-2.0\rho}}{\rho}$ to avoid large additive offsets (see \hyperref[fourth:app]{Connecting the entropy and the biological reliability cost}--Eq.\ref{eq:csrho} for details).
    \textbf{c.} The synaptic variability as a function of the posterior variance for different values of $c$: $[0.112, 0.2, 0.356]$ (3 DP). As $c$ increases (lighter blues) we penalise reliability more, and hence the optimised synaptic noise variability increases. (Here we fixed $\rho=1/2$ across different settings for $c$)}
    \label{fig:geometric}
\end{figure}

To assess the connection between Bayesian posteriors and energy efficient variability in more depth, we estimated and plotted the posterior variance against the optimised synaptic variability (Fig.~\ref{fig:geometric}a) \add{(see \hyperref[sec:methods]{Methods})}.
We considered our four different biophysical mechanisms (values for $\rho$; Fig.~\ref{fig:geometric}a, columns), and values for $c$ (Fig.~\ref{fig:geometric}a, rows).
In all cases, there was a clear correlation between the posterior and the optimised variability: synapses with larger posterior variance also had large optimised variance.
To further assess this connection, we used the relation between the Hessian and posterior variance given by Eq.~\ref{eq:identities}c and the analytic result given in \hyperref[fifth:app]{Appendix - Analytic predictions for $\sigma_i$} 
to plot the relationships between $\sigma_i$ and the posterior variability, $\sigma_\text{post}$ as a function of $\rho$ (Fig.~\ref{fig:geometric}b;) and as a function of $c$ (Fig.~\ref{fig:geometric}c).
Again, these plots show a clear correlation between synapse variance and posterior variance; though the relationship is far from perfect.
For a perfect relationship, we would expect the lines in Fig.~\ref{fig:geometric}bc to all lie along the diagonal \add{with slope equal to one}.
In contrast, these lines actually have a slope smaller than one, indicating that optimised variability is less heterogeneous than posterior variance (Fig.~\ref{fig:geometric}bc).
\add{Interestingly, the slope increases towards one as the associated $\rho$ is decreased, this suggests that synapse variability best approximates the posterior when $\rho$ is small.} 

\add{This strong, but not perfect, connection between the patterns of variability in Bayesian inference and energy-efficient networks motivated us to seek a formal connection between Bayesian and efficient synapses.}
As such, in the Appendix, we derive a theoretical connection between our overall performance cost and Bayesian inference  
\add{(see \hyperref[fourth:app]{Appendix - Energy efficient noise and variational-Bayes for neural network weights}). Moreover, this connection is subsequently used to provide an explanation for why synapse variability aligns closer to posterior variance for small $\rho$ (see Eq.~\ref{eq:smallrho})}.
Specifically, variational inference, a well-known procedure for performing (approximate) Bayesian inference in NNs \citep{hinton93keeping, graves2011practical, blundell2015weight}.
Variational inference optimises the ``evidence lower bound objective'' (ELBO) \citep{barber1998ensemble, jordan1999introduction, blei2017variational}, which surprisingly turns out to resemble our performance cost.
Specifically, the ELBO includes a term which encourages the entropy of the approximating posterior distribution (which could be interpreted as our noise distribution) to be larger.
This resembles a reliability cost, as our reliability costs also encourage the noise distribution to be larger.
Critically, the biological power-law reliability cost has a different form from the ideal, entropic reliability cost.
However, we are able to derive a formal relationship: the biological power-law reliability costs bounds the ideal entropic reliability cost. 
Remarkably, this implies that our overall cost (Eq.~\ref{eq:wordobjective}) bounds the ELBO, so reducing our cost (Eq.~\ref{eq:wordobjective}) tightens the ELBO bound and gives an improved guarantee on the quality of Bayesian inference.

\section{Discussion}
\label{sec:dis}

Comparing the brain's computational roles with associated energetic costs provides a useful means for deducing properties of efficient neurophysiology. 
Here, we applied this approach to PSP variability.
We began by looking at the biophysical mechanisms of synaptic transmission, and how the energy costs for transmission might vary with synaptic reliability.
We modified a standard ANN to incorporate unreliable synapses and trained this on a classification task using an objective that combined classification accuracy and an energetic cost on reliability. 
This led to a performance-reliability cost tradeoff and heterogeneous patterns of synapse variability that correlated with input rate and learning rate.
We noted that these patterns of variability have been previously observed in data (see Fig.~\ref{fig:exp}).
Remarkably, these are also the patterns of variability predicted by Bayesian synapses \citep{aitchison2021synaptic} (i.e.\ when distributions over synaptic weights correspond with the Bayesian posterior).
Finally, we showed empirical and formal connections between the synaptic variability implied by Bayesian synapses and our performance-reliability cost tradeoff.

The reliability cost in terms of the synaptic variability (Eq.~\ref{eq:reliabilityCost}) is a critical component of the numerical experiments we present here.
While the precise form of the cost is inevitably uncertain, we attempted to mitigate the uncertainty by considering a wide range of functional forms for the reliability cost.
In particular, we considered four biophysical mechanisms, corresponding to four power-law exponents, ($\rho = \tfrac{1}{2}, \tfrac{2}{3}, \tfrac{4}{3}, 2$).
Moreover, these different power-law costs already cover a reasonably wide-range of potential penalties and we would expect the results to hold for many other forms of reliability cost as the intuition behind the results ultimately relies merely on there being \textit{some} penalty for increasing reliability. 

The biophysical cost also includes a multiplicative factor, $c$, which sets the magnitude of the reliability cost. 
In fact, the patterns of variability exhibited in Fig.~\ref{fig:optimising} are preserved as $c$ is changed: this was demonstrated for values of $c$ which are ten times larger and ten times smaller, \hyperref[sixth:app]{Supplementary - Appendix 6--Fig.~\ref{fig:varying_c}}.
This multiplicative factor should be understood as being determined by the properties of the physics and chemistry underpinning synaptic dynamics, for example it could represent the quantity of ATP required by the metabolic costs of synaptic transmission (although this factor could vary e.g.\ in different cell types).

Our artificial neural networks used backpropagation to optimise the mean and variance of synaptic weights.
While there are a number of schemes by which biological circuits might implement backpropagation \citep{whittington2017approximation, sacramento2018dendritic, richards2019dendritic}, it is not yet clear whether backpropagation is implemented by the brain (see \citet{lillicrap2020backpropagation} for a review on the plausibility of propagation in the brain). 
Regardless, backpropagation is merely the route we used in our ANN setting to reach an energy efficient configuration. 
The patterns we have observed are characteristic of an energy-efficient network and therefore should not depend on the learning rule that the brain uses to achieve energy-efficiency.

Our results in ANNs used MNIST classification as an example of a task; this may appear somewhat artificial, but all brain areas ultimately do have a task: to maximise fitness (or reward as a proxy for fitness).
Moreover, our results all ultimately arise from trading off biophysical reliability costs against the fact that if a synapse is important to performing a task, then variability in that synapse substantially impairs performance.
Of course performance, in different brain areas, might mean reward, fitness or some other measures.
In contrast, if a synapse is unimportant, variability in that synapse impairs performance less.
In all tasks there will be some synapses that are more, and some synapses that are less important, and our task, while relatively straightforward, captures this important property.


Our results have important implications for the understanding of Bayesian inference in synapses. 
In particular, we show that energy efficiency considerations give rise to two phenomena that are consistent with predictions outlined in previous work on Bayesian synapses \citep{aitchison2021synaptic}. 
First, that normalised variability decreases for synapses with higher presynaptic firing rates. 
Second, that synaptic plasticity is higher for synapses with higher variability. 

Specifically, these findings suggest that synapses connect their uncertainty in the value of the optimal synaptic weight \citep[see][for details]{aitchison2021synaptic} to variability.
This is in essence a synaptic variant of the ``sampling hypothesis''. 
Under the sampling hypothesis, neural activity is believed to represent a potential state of the world, and variability is believed to represent uncertainty \citep{hoyer2002interpreting,knill2004bayesian, ma2006bayesian,fiser2010statistically,berkes2011spontaneous,orban2016neural,aitchison2016hamiltonian,haefner2016perceptual,lange2017characterizing,shivkumar2018probabilistic,bondy2018feedback,echeveste2020cortical,festa2021neuronal,lange2021confirmation,lange2022task}.
This variability in neural activity, representing uncertainty in the state of the world, can then be read out by downstream circuits to inform behaviour.
Here, we showed that a connection between synaptic uncertainty and variability can emerge simply as a consequence of maximising energy efficiency.
This suggest that Bayesian synapses may emerge without any necessity for specific synaptic biophysical implementations of Bayesian inference.

%

Importantly though, while the brain might use synaptic noise for Bayesian computation, these results are also consistent with an alternative interpretation: that the brain is not Bayesian, it just looks Bayesian because it is energy efficient.
To distinguish between these two interpretations, we ultimately need to know whether downstream brain areas exploit or ignore information about uncertainty that arises from synaptic variability.

\section{Materials and methods}
\label{sec:methods}
The ANN simulations were run in PyTorch with feedforward, fully-connected neural networks with two hidden layers of width 100. The input dimension of 784 corresponded to the number of pixels in the greyscale MNIST images of handwritten digits, while the output dimension of ten corresponded to the number of classes. We used the reparameterisation trick to backpropagate with respect to the mean and variance of the weights, in particular, we set $w_i = \mu_i + \sigma_i\xi$ where $\xi \sim \text{Normal}(0, 1)$ \citep{kingma2015variational}.
MNIST classification was learned through optimisation of Gaussian parameters with respect to a cross-entropy loss in addition to reliability costs using minibatch gradient descent under Adam optimisation with a minibatch size of 20. 
To prevent negative values for the $\sigma$s, they were re-parameterised using a softplus function with argument $\phi_i$, with $\sigma_i=\text{softplus}(\phi_i)$. The base learning rate in Eq.\ref{eq:adaptive_learning_rate} is $\eta_\text{base}= 5\times 10^{-4}$. The $\mu_i$s were initialised homogeneously across the network from $\text{Uniform}(-0.1, 0.1)$ and the $\sigma_i$s were initialised homogeneously across the network at $10^{-4}$. Hyperparameters were chosen via grid search on the validation dataset to enable smooth learning, high performance and rapid convergence. In the objective $\Lbi$ used to train our simulations, we also add an L1 regularisation term over synaptic weights, $\lambda|\mu|_1$, where $\lambda=10^{-4}$. 

Plots in Fig.~\ref{fig:accuracy} present mappings from hyperparameter $c$, to accuracy and $\sigma$. A different neural network was trained for each $c$, after 50 training epochs the average $\sigma$ across synapses was computed, and accuracy was evaluated on the test dataset.  Plots in Fig.~\ref{fig:performance} present mappings of this $\sigma$  against accuracy and reliability cost. The reliability cost was computed using fixed $s=1$ (see Appendix~\ref{eq:csrho}). 


To compute the Hessian in Fig.~\ref{fig:optimising} and elsewhere, we used the empirical Fisher information approximation \citep{fisher1922mathematical}, $H\approx g^2$. This was evaluated by taking the average $g^2$ at $w^*=\mu$ over ten epochs after full training for 50 epochs. The average learning rate $\gamma|g|^{-1}$ and the average input rate $|x|$ were also evaluated over ten epochs following training. The data presented illustrate these variables with regard to the weights of the second hidden layer. We set hyperparameter $s=0.001$ (see Eq.\ref{eq:csrho}) in these simulations.


\replace{For geometric comparisons between the distribution over synapses and the Bayesian posterior presented in Fig.~\ref{fig:geometric} we used the analytic results in \hyperref[fifth:app]{Appendix - Analytic predictions for $\sigma_i$}.}{
To estimate the posterior used in Fig.~\ref{fig:geometric} we optimised a factorised Gaussian approximation to the posterior over weights using variational inference and Bayes by back-propagation \citep{blundell2015weight}. We then took $\sigma_{post}$ from two optimised weights. For the variance and slope comparisons between Bayesian and efficient synapses in Fig.~\ref{fig:geometric} we used the analytic results in \hyperref[fifth:app]{Appendix - Analytic predictions for $\sigma_i$}.}

Source code used for simulations available at \href{https://github.com/JamesMalkin/EfficientBayes.git}{github.com/JamesMalkin/EfficientBayes}

\section{Acknowledgments}
We are grateful to Dr Stewart whose philanthropy supported GPU compute used in this project. JM was funded by the Engineering and Physical Sciences Research Council (2482786). COD was funded by the Leverhulme Trust (RPG-2019-229) and Biotechnology and Biological Sciences Research Council (BB/W001845/1). CH is supported by the Leverhulme Trust (RF-2021-533). 

\bibliography{references}

\begin{thebibliography}{82}
\providecommand{\natexlab}[1]{#1}
\providecommand{\urlprefix}{}
\providecommand{\doiprefix}{doi: }

\bibitem[{Aitchison(2020)Aitchison, Laurence}]{aitchison2020bayesian}
\textbf{\color{eLifeMediumGrey} Aitchison L}.
\newblock Bayesian filtering unifies adaptive and non-adaptive neural network optimization methods.
\newblock Advances in Neural Information Processing Systems.  2020; 33:18173--18182.

\bibitem[{Aitchison et~al.(2021)Aitchison, Laurence and Jegminat, Jannes and Menendez, Jorge Aurelio and Pfister, Jean-Pascal and Pouget, Alexandre and Latham, Peter E}]{aitchison2021synaptic}
\textbf{\color{eLifeMediumGrey} Aitchison L}, Jegminat J, Menendez JA, Pfister JP, Pouget A, Latham PE.
\newblock Synaptic plasticity as Bayesian inference.
\newblock Nature Neuroscience.  2021; 24(4):565--571.

\bibitem[{Aitchison and Lengyel(2016)Aitchison, Laurence and Lengyel, M{\'a}t{\'e}}]{aitchison2016hamiltonian}
\textbf{\color{eLifeMediumGrey} Aitchison L}, Lengyel M.
\newblock The Hamiltonian brain: Efficient probabilistic inference with excitatory-inhibitory neural circuit dynamics.
\newblock PLoS computational biology.  2016; 12(12):e1005186.

\bibitem[{Attwell and Laughlin(2001)Attwell, David and Laughlin, Simon B}]{attwell2001energy}
\textbf{\color{eLifeMediumGrey} Attwell D}, Laughlin SB.
\newblock An energy budget for signaling in the grey matter of the brain.
\newblock Journal of Cerebral Blood Flow \& Metabolism.  2001; 21(10):1133--1145.

\bibitem[{Barber and Bishop(1998)Barber, David and Bishop, Christopher M}]{barber1998ensemble}
\textbf{\color{eLifeMediumGrey} Barber D}, Bishop CM.
\newblock Ensemble learning in Bayesian neural networks.
\newblock Nato ASI Series F Computer and Systems Sciences.  1998; 168:215--238.

\bibitem[{Bellingham et~al.(1998)Bellingham, Mark C and Lim, Rebecca and Walmsley, Bruce}]{bellingham1998developmental}
\textbf{\color{eLifeMediumGrey} Bellingham MC}, Lim R, Walmsley B.
\newblock Developmental changes in EPSC quantal size and quantal content at a central glutamatergic synapse in rat.
\newblock The Journal of Physiology.  1998; 511(3):861--869.

\bibitem[{Berkes et~al.(2011)Berkes, Pietro and Orb{\'a}n, Gerg{\H{o}} and Lengyel, M{\'a}t{\'e} and Fiser, J{\'o}zsef}]{berkes2011spontaneous}
\textbf{\color{eLifeMediumGrey} Berkes P}, Orb{\'a}n G, Lengyel M, Fiser J.
\newblock Spontaneous cortical activity reveals hallmarks of an optimal internal model of the environment.
\newblock Science.  2011; 331(6013):83--87.

\bibitem[{Blei et~al.(2017)Blei, David M and Kucukelbir, Alp and McAuliffe, Jon D}]{blei2017variational}
\textbf{\color{eLifeMediumGrey} Blei DM}, Kucukelbir A, McAuliffe JD.
\newblock Variational inference: A review for statisticians.
\newblock Journal of the American statistical Association.  2017; 112(518):859--877.

\bibitem[{Blundell et~al.(2015)Blundell, Charles and Cornebise, Julien and Kavukcuoglu, Koray and Wierstra, Daan}]{blundell2015weight}
\textbf{\color{eLifeMediumGrey} Blundell C}, Cornebise J, Kavukcuoglu K, Wierstra D.
\newblock Weight uncertainty in neural network.
\newblock In: \emph{International conference on machine learning} PMLR; 2015. p. 1613--1622.

\bibitem[{Bondy et~al.(2018)Bondy, Adrian G and Haefner, Ralf M and Cumming, Bruce G}]{bondy2018feedback}
\textbf{\color{eLifeMediumGrey} Bondy AG}, Haefner RM, Cumming BG.
\newblock Feedback determines the structure of correlated variability in primary visual cortex.
\newblock Nature neuroscience.  2018; 21(4):598--606.

\bibitem[{Boyd and Martin(1956)Boyd, IA and Martin, AR}]{boyd1956end}
\textbf{\color{eLifeMediumGrey} Boyd I}, Martin A.
\newblock The end-plate potential in mammalian muscle.
\newblock The Journal of physiology.  1956; 132(1):74.

\bibitem[{Branco and Staras(2009)Branco, Tiago and Staras, Kevin}]{branco2009probability}
\textbf{\color{eLifeMediumGrey} Branco T}, Staras K.
\newblock The probability of neurotransmitter release: variability and feedback control at single synapses.
\newblock Nature Reviews Neuroscience.  2009; 10(5):373--383.

\bibitem[{Bridgman(1999)Bridgman, PC}]{bridgman1999myosin}
\textbf{\color{eLifeMediumGrey} Bridgman P}.
\newblock Myosin Va movements in normal and dilute-lethal axons provide support for a dual filament motor complex.
\newblock The Journal of Cell Biology.  1999; 146(5):1045--1060.

\bibitem[{Brock et~al.(2020)Brock, Jennifer A and Thomazeau, Aurore and Watanabe, Airi and Li, Sally Si Ying and Sj{\"o}str{\"o}m, P Jesper}]{brock2020practical}
\textbf{\color{eLifeMediumGrey} Brock JA}, Thomazeau A, Watanabe A, Li SSY, Sj{\"o}str{\"o}m PJ.
\newblock A practical guide to using CV analysis for determining the locus of synaptic plasticity.
\newblock Frontiers in Synaptic Neuroscience.  2020; 12:11.

\bibitem[{Chenouard et~al.(2020)Chenouard, Nicolas and Xuan, Feng and Tsien, Richard W}]{chenouard2020synaptic}
\textbf{\color{eLifeMediumGrey} Chenouard N}, Xuan F, Tsien RW.
\newblock Synaptic vesicle traffic is supported by transient actin filaments and regulated by PKA and NO.
\newblock Nature communications.  2020; 11(1):5318.

\bibitem[{Cingolani and Goda(2008)Cingolani, Lorenzo A and Goda, Yukiko}]{cingolani2008actin}
\textbf{\color{eLifeMediumGrey} Cingolani LA}, Goda Y.
\newblock Actin in action: the interplay between the actin cytoskeleton and synaptic efficacy.
\newblock Nature Reviews Neuroscience.  2008; 9(5):344--356.

\bibitem[{Davis and M{\"u}ller(2015)Davis, Graeme W and M{\"u}ller, Martin}]{davis2015homeostatic}
\textbf{\color{eLifeMediumGrey} Davis GW}, M{\"u}ller M.
\newblock Homeostatic control of presynaptic neurotransmitter release.
\newblock Annual review of physiology.  2015; 77:251--270.

\bibitem[{Dobrunz and Stevens(1997)Dobrunz, Lynn E and Stevens, Charles F}]{dobrunz1997heterogeneity}
\textbf{\color{eLifeMediumGrey} Dobrunz LE}, Stevens CF.
\newblock Heterogeneity of release probability, facilitation, and depletion at central synapses.
\newblock Neuron.  1997; 18(6):995--1008.

\bibitem[{Echeveste et~al.(2020)Echeveste, Rodrigo and Aitchison, Laurence and Hennequin, Guillaume and Lengyel, M{\'a}t{\'e}}]{echeveste2020cortical}
\textbf{\color{eLifeMediumGrey} Echeveste R}, Aitchison L, Hennequin G, Lengyel M.
\newblock Cortical-like dynamics in recurrent circuits optimized for sampling-based probabilistic inference.
\newblock Nature neuroscience.  2020; 23(9):1138--1149.

\bibitem[{Engl and Attwell(2015)Engl, Elisabeth and Attwell, David}]{engl2015non}
\textbf{\color{eLifeMediumGrey} Engl E}, Attwell D.
\newblock Non-signalling energy use in the brain.
\newblock The Journal of physiology.  2015; 593(16):3417--3429.

\bibitem[{Festa et~al.(2021)Festa, Dylan and Aschner, Amir and Davila, Aida and Kohn, Adam and Coen-Cagli, Ruben}]{festa2021neuronal}
\textbf{\color{eLifeMediumGrey} Festa D}, Aschner A, Davila A, Kohn A, Coen-Cagli R.
\newblock Neuronal variability reflects probabilistic inference tuned to natural image statistics.
\newblock Nature communications.  2021; 12(1):3635.

\bibitem[{Fiser et~al.(2010)Fiser, J{\'o}zsef and Berkes, Pietro and Orb{\'a}n, Gerg{\H{o}} and Lengyel, M{\'a}t{\'e}}]{fiser2010statistically}
\textbf{\color{eLifeMediumGrey} Fiser J}, Berkes P, Orb{\'a}n G, Lengyel M.
\newblock Statistically optimal perception and learning: from behavior to neural representations.
\newblock Trends in cognitive sciences.  2010; 14(3):119--130.

\bibitem[{Fisher(1922)Fisher, Ronald A}]{fisher1922mathematical}
\textbf{\color{eLifeMediumGrey} Fisher RA}.
\newblock On the mathematical foundations of theoretical statistics.
\newblock Philosophical transactions of the Royal Society of London Series A, containing papers of a mathematical or physical character.  1922; 222(594-604):309--368.

\bibitem[{Forti et~al.(1997)Forti, Lia and Bossi, Mario and Bergamaschi, Andrea and Villa, Antonello and Malgaroli, Antonio}]{forti1997loose}
\textbf{\color{eLifeMediumGrey} Forti L}, Bossi M, Bergamaschi A, Villa A, Malgaroli A.
\newblock Loose-patch recordings of single quanta at individual hippocampal synapses.
\newblock Nature.  1997; 388(6645):874--878.

\bibitem[{Fukushima(1975)Fukushima, Kunihiko}]{fukushima1975cognitron}
\textbf{\color{eLifeMediumGrey} Fukushima K}.
\newblock Cognitron: A self-organizing multilayered neural network.
\newblock Biological cybernetics.  1975; 20(3-4):121--136.

\bibitem[{Gentile et~al.(2022)Gentile, Juliana E and Carrizales, Melissa G and Koleske, Anthony J}]{gentile2022control}
\textbf{\color{eLifeMediumGrey} Gentile JE}, Carrizales MG, Koleske AJ.
\newblock Control of synapse structure and function by actin and its regulators.
\newblock Cells.  2022; 11(4):603.

\bibitem[{Goldman(2004)Goldman, Mark S}]{goldman2004enhancement}
\textbf{\color{eLifeMediumGrey} Goldman MS}.
\newblock Enhancement of information transmission efficiency by synaptic failures.
\newblock Neural Computation.  2004; 16(6):1137--1162.

\bibitem[{Gramlich and Klyachko(2017)Gramlich, Michael W and Klyachko, Vitaly A}]{gramlich2017actin}
\textbf{\color{eLifeMediumGrey} Gramlich MW}, Klyachko VA.
\newblock Actin/Myosin-V-and activity-dependent inter-synaptic vesicle exchange in central neurons.
\newblock Cell Reports.  2017; 18(9):2096--2104.

\bibitem[{Graves(2011)Graves, Alex}]{graves2011practical}
\textbf{\color{eLifeMediumGrey} Graves A}.
\newblock Practical variational inference for neural networks.
\newblock Advances in Neural Information Processing Systems.  2011; 24.

\bibitem[{Haefner et~al.(2016)Haefner, Ralf M and Berkes, Pietro and Fiser, J{\'o}zsef}]{haefner2016perceptual}
\textbf{\color{eLifeMediumGrey} Haefner RM}, Berkes P, Fiser J.
\newblock Perceptual decision-making as probabilistic inference by neural sampling.
\newblock Neuron.  2016; 90(3):649--660.

\bibitem[{Harris et~al.(2012)Harris, Julia J and Jolivet, Renaud and Attwell, David}]{harris2012synaptic}
\textbf{\color{eLifeMediumGrey} Harris JJ}, Jolivet R, Attwell D.
\newblock Synaptic energy use and supply.
\newblock Neuron.  2012; 75(5):762--777.

\bibitem[{Harris et~al.(2019)Harris, Julia Jade and Engl, Elisabeth and Attwell, David and Jolivet, Renaud Blaise}]{harris2019energy}
\textbf{\color{eLifeMediumGrey} Harris JJ}, Engl E, Attwell D, Jolivet RB.
\newblock Energy-efficient information transfer at thalamocortical synapses.
\newblock PLoS computational biology.  2019; 15(8):e1007226.

\bibitem[{Heidelberger et~al.(1994)Heidelberger, Ruth and Heinemann, Christian and Neher, Erwin and Matthews, Gary}]{heidelberger1994calcium}
\textbf{\color{eLifeMediumGrey} Heidelberger R}, Heinemann C, Neher E, Matthews G.
\newblock Calcium dependence of the rate of exocytosis in a synaptic terminal.
\newblock Nature.  1994; 371(6497):513--515.

\bibitem[{Hinton and van Camp(1993)Hinton, GE and van Camp, Drew}]{hinton93keeping}
\textbf{\color{eLifeMediumGrey} Hinton G}, van Camp D.
\newblock Keeping neural networks simple by minimising the description length of weights. 1993.
\newblock In: \emph{Proceedings of COLT-93}; 1993. p. 5--13.

\bibitem[{Holler et~al.(2021)Holler, Simone and K{\"o}stinger, German and Martin, Kevan AC and Schuhknecht, Gregor FP and Stratford, Ken J}]{holler2021structure}
\textbf{\color{eLifeMediumGrey} Holler S}, K{\"o}stinger G, Martin KA, Schuhknecht GF, Stratford KJ.
\newblock Structure and function of a neocortical synapse.
\newblock Nature.  2021; 591(7848):111--116.

\bibitem[{Hoyer and Hyv{\"a}rinen(2002)Hoyer, Patrik and Hyv{\"a}rinen, Aapo}]{hoyer2002interpreting}
\textbf{\color{eLifeMediumGrey} Hoyer P}, Hyv{\"a}rinen A.
\newblock Interpreting neural response variability as Monte Carlo sampling of the posterior.
\newblock Advances in neural information processing systems.  2002; 15.

\bibitem[{Jegminat et~al.(2022)Jegminat, Jannes and Surace, Simone Carlo and Pfister, Jean-Pascal}]{jegminat2022learning}
\textbf{\color{eLifeMediumGrey} Jegminat J}, Surace SC, Pfister JP.
\newblock Learning as filtering: Implications for spike-based plasticity.
\newblock PLoS computational biology.  2022; 18(2):e1009721.

\bibitem[{Jordan et~al.(1999)Jordan, Michael I and Ghahramani, Zoubin and Jaakkola, Tommi S and Saul, Lawrence K}]{jordan1999introduction}
\textbf{\color{eLifeMediumGrey} Jordan MI}, Ghahramani Z, Jaakkola TS, Saul LK.
\newblock An introduction to variational methods for graphical models.
\newblock Machine learning.  1999; 37:183--233.

\bibitem[{Karbowski(2019)Karbowski, Jan}]{karbowski2019metabolic}
\textbf{\color{eLifeMediumGrey} Karbowski J}.
\newblock Metabolic constraints on synaptic learning and memory.
\newblock Journal of Neurophysiology.  2019; 122(4):1473--1490.

\bibitem[{Karunanithi et~al.(2002)Karunanithi, Shanker and Marin, Leo and Wong, Kar and Atwood, Harold L}]{karunanithi2002quantal}
\textbf{\color{eLifeMediumGrey} Karunanithi S}, Marin L, Wong K, Atwood HL.
\newblock Quantal size and variation determined by vesicle size in normal and mutant Drosophila glutamatergic synapses.
\newblock Journal of Neuroscience.  2002; 22(23):10267--10276.

\bibitem[{Katz and Miledi(1965)Katz, Bernard and Miledi, Ricardo}]{katz1965measurement}
\textbf{\color{eLifeMediumGrey} Katz B}, Miledi R.
\newblock The measurement of synaptic delay, and the time course of acetylcholine release at the neuromuscular junction.
\newblock Proceedings of the Royal Society of London Series B Biological Sciences.  1965; 161(985):483--495.

\bibitem[{Khan et~al.(2018)Khan, Mohammad and Nielsen, Didrik and Tangkaratt, Voot and Lin, Wu and Gal, Yarin and Srivastava, Akash}]{khan2018fast}
\textbf{\color{eLifeMediumGrey} Khan M}, Nielsen D, Tangkaratt V, Lin W, Gal Y, Srivastava A.
\newblock Fast and scalable bayesian deep learning by weight-perturbation in adam.
\newblock In: \emph{International conference on machine learning} PMLR; 2018. p. 2611--2620.

\bibitem[{Kingma and Ba(2014)Kingma, Diederik P and Ba, Jimmy}]{kingma2014adam}
\textbf{\color{eLifeMediumGrey} Kingma DP}, Ba J.
\newblock Adam: A method for stochastic optimization.
\newblock arXiv preprint arXiv:14126980.  2014; .

\bibitem[{Kingma et~al.(2015)Kingma, Durk P and Salimans, Tim and Welling, Max}]{kingma2015variational}
\textbf{\color{eLifeMediumGrey} Kingma DP}, Salimans T, Welling M.
\newblock Variational dropout and the local reparameterization trick.
\newblock Advances in Neural Information Processing Systems.  2015; 28.

\bibitem[{Knill and Pouget(2004)Knill, David C and Pouget, Alexandre}]{knill2004bayesian}
\textbf{\color{eLifeMediumGrey} Knill DC}, Pouget A.
\newblock The Bayesian brain: the role of uncertainty in neural coding and computation.
\newblock TRENDS in Neurosciences.  2004; 27(12):712--719.

\bibitem[{Ko et~al.(2013)Ko, Ho and Cossell, Lee and Baragli, Chiara and Antolik, Jan and Clopath, Claudia and Hofer, Sonja B and Mrsic-Flogel, Thomas D}]{ko2013emergence}
\textbf{\color{eLifeMediumGrey} Ko H}, Cossell L, Baragli C, Antolik J, Clopath C, Hofer SB, Mrsic-Flogel TD.
\newblock The emergence of functional microcircuits in visual cortex.
\newblock Nature.  2013; 496(7443):96--100.

\bibitem[{Lange et~al.(2021)Lange, Richard D and Chattoraj, Ankani and Beck, Jeffrey M and Yates, Jacob L and Haefner, Ralf M}]{lange2021confirmation}
\textbf{\color{eLifeMediumGrey} Lange RD}, Chattoraj A, Beck JM, Yates JL, Haefner RM.
\newblock A confirmation bias in perceptual decision-making due to hierarchical approximate inference.
\newblock PLoS Computational Biology.  2021; 17(11):e1009517.

\bibitem[{Lange and Haefner(2017)Lange, Richard D and Haefner, Ralf M}]{lange2017characterizing}
\textbf{\color{eLifeMediumGrey} Lange RD}, Haefner RM.
\newblock Characterizing and interpreting the influence of internal variables on sensory activity.
\newblock Current opinion in neurobiology.  2017; 46:84--89.

\bibitem[{Lange and Haefner(2022)Lange, Richard D and Haefner, Ralf M}]{lange2022task}
\textbf{\color{eLifeMediumGrey} Lange RD}, Haefner RM.
\newblock Task-induced neural covariability as a signature of approximate Bayesian learning and inference.
\newblock PLoS computational biology.  2022; 18(3):e1009557.

\bibitem[{Laughlin et~al.(1998)Laughlin, Simon B and de Ruyter van Steveninck, Rob R and Anderson, John C}]{laughlin1998metabolic}
\textbf{\color{eLifeMediumGrey} Laughlin SB}, de~Ruyter~van Steveninck RR, Anderson JC.
\newblock The metabolic cost of neural information.
\newblock Nature Neuroscience.  1998; 1(1):36--41.

\bibitem[{LeCun et~al.(2002)LeCun, Yann and Bottou, L{\'e}on and Orr, Genevieve B and M{\"u}ller, Klaus-Robert}]{lecun2002efficient}
\textbf{\color{eLifeMediumGrey} LeCun Y}, Bottou L, Orr GB, M{\"u}ller KR.
\newblock Efficient backprop.
\newblock In: \emph{Neural networks: Tricks of the trade} Springer; 2002.p. 9--50.

\bibitem[{Levy and Baxter(2002)Levy, William B and Baxter, Robert A}]{levy2002energy}
\textbf{\color{eLifeMediumGrey} Levy WB}, Baxter RA.
\newblock Energy-efficient neuronal computation via quantal synaptic failures.
\newblock Journal of Neuroscience.  2002; 22(11):4746--4755.

\bibitem[{Lillicrap et~al.(2020)Lillicrap, Timothy P and Santoro, Adam and Marris, Luke and Akerman, Colin J and Hinton, Geoffrey}]{lillicrap2020backpropagation}
\textbf{\color{eLifeMediumGrey} Lillicrap TP}, Santoro A, Marris L, Akerman CJ, Hinton G.
\newblock Backpropagation and the brain.
\newblock Nature Reviews Neuroscience.  2020; 21(6):335--346.

\bibitem[{Lisman and Harris(1993)Lisman, John E and Harris, Kristen M}]{lisman1993quantal}
\textbf{\color{eLifeMediumGrey} Lisman JE}, Harris KM.
\newblock Quantal analysis and synaptic anatomy—integrating two views of hippocampal plasticity.
\newblock Trends in Neurosciences.  1993; 16(4):141--147.

\bibitem[{Ma et~al.(2006)Ma, Wei Ji and Beck, Jeffrey M and Latham, Peter E and Pouget, Alexandre}]{ma2006bayesian}
\textbf{\color{eLifeMediumGrey} Ma WJ}, Beck JM, Latham PE, Pouget A.
\newblock Bayesian inference with probabilistic population codes.
\newblock Nature neuroscience.  2006; 9(11):1432--1438.

\bibitem[{MacKay(1992{\natexlab{a}})MacKay, David JC}]{mackay1992evidence}
\textbf{\color{eLifeMediumGrey} MacKay DJ}.
\newblock The evidence framework applied to classification networks.
\newblock Neural computation.  1992; 4(5):720--736.

\bibitem[{MacKay(1992{\natexlab{b}})MacKay, David JC}]{mackay1992practical}
\textbf{\color{eLifeMediumGrey} MacKay DJ}.
\newblock A practical Bayesian framework for backpropagation networks.
\newblock Neural Computation.  1992; 4(3):448--472.

\bibitem[{Martens(2020)Martens, James}]{martens2020new}
\textbf{\color{eLifeMediumGrey} Martens J}.
\newblock New insights and perspectives on the natural gradient method.
\newblock The Journal of Machine Learning Research.  2020; 21(1):5776--5851.

\bibitem[{Murphy(2012)Murphy, Kevin P}]{murphy2012machine}
\textbf{\color{eLifeMediumGrey} Murphy KP}.
\newblock Machine learning: a probabilistic perspective.
\newblock MIT press; 2012.

\bibitem[{Murthy et~al.(1997)Murthy, Venkatesh N and Sejnowski, Terrence J and Stevens, Charles F}]{murthy1997heterogeneous}
\textbf{\color{eLifeMediumGrey} Murthy VN}, Sejnowski TJ, Stevens CF.
\newblock Heterogeneous release properties of visualized individual hippocampal synapses.
\newblock Neuron.  1997; 18(4):599--612.

\bibitem[{Orb{\'a}n et~al.(2016)Orb{\'a}n, Gerg{\H{o}} and Berkes, Pietro and Fiser, J{\'o}zsef and Lengyel, M{\'a}t{\'e}}]{orban2016neural}
\textbf{\color{eLifeMediumGrey} Orb{\'a}n G}, Berkes P, Fiser J, Lengyel M.
\newblock Neural variability and sampling-based probabilistic representations in the visual cortex.
\newblock Neuron.  2016; 92(2):530--543.

\bibitem[{Paulsen and Heggelund(1996)Paulsen, O and Heggelund, P}]{paulsen1996quantal}
\textbf{\color{eLifeMediumGrey} Paulsen O}, Heggelund P.
\newblock Quantal properties of spontaneous EPSCs in neurones of the guinea-pig dorsal lateral geniculate nucleus.
\newblock The Journal of Physiology.  1996; 496(3):759--772.

\bibitem[{Paulsen and Heggelund(1994)Paulsen, Ole and Heggelund, Paul}]{paulsen1994quantal}
\textbf{\color{eLifeMediumGrey} Paulsen O}, Heggelund P.
\newblock The quantal size at retinogeniculate synapses determined from spontaneous and evoked EPSCs in guinea-pig thalamic slices.
\newblock The Journal of Physiology.  1994; 480(3):505--511.

\bibitem[{Pulido and Ryan(2021)Pulido, Camila and Ryan, Timothy A}]{pulido2021synaptic}
\textbf{\color{eLifeMediumGrey} Pulido C}, Ryan TA.
\newblock Synaptic vesicle pools are a major hidden resting metabolic burden of nerve terminals.
\newblock Science Advances.  2021; 7(49):eabi9027.

\bibitem[{Purdon et~al.(2002)Purdon, AD and Rosenberger, TA and Shetty, HU and Rapoport, SI}]{purdon2002energy}
\textbf{\color{eLifeMediumGrey} Purdon A}, Rosenberger T, Shetty H, Rapoport S.
\newblock Energy consumption by phospholipid metabolism in mammalian brain.
\newblock Neurochemical Research.  2002; 27(12):1641--1647.

\bibitem[{Raghavachari and Lisman(2004)Raghavachari, Sridhar and Lisman, John E}]{raghavachari2004properties}
\textbf{\color{eLifeMediumGrey} Raghavachari S}, Lisman JE.
\newblock Properties of quantal transmission at CA1 synapses.
\newblock Journal of neurophysiology.  2004; 92(4):2456--2467.

\bibitem[{Richards and Lillicrap(2019)Richards, Blake A and Lillicrap, Timothy P}]{richards2019dendritic}
\textbf{\color{eLifeMediumGrey} Richards BA}, Lillicrap TP.
\newblock Dendritic solutions to the credit assignment problem.
\newblock Current opinion in neurobiology.  2019; 54:28--36.

\bibitem[{Rosset and Zhu(2006)Rosset, Saharon and Zhu, Ji}]{rosset2006sparse}
\textbf{\color{eLifeMediumGrey} Rosset S}, Zhu J.
\newblock Sparse, Flexible and Efficient Modeling using L 1 Regularization.
\newblock Feature Extraction: Foundations and Applications.  2006; p. 375--394.

\bibitem[{Sacramento et~al.(2018)Sacramento, Jo{\~a}o and Ponte Costa, Rui and Bengio, Yoshua and Senn, Walter}]{sacramento2018dendritic}
\textbf{\color{eLifeMediumGrey} Sacramento J}, Ponte~Costa R, Bengio Y, Senn W.
\newblock Dendritic cortical microcircuits approximate the backpropagation algorithm.
\newblock Advances in neural information processing systems.  2018; 31.

\bibitem[{Sacramento et~al.(2015)Sacramento, Jo{\~a}o and Wichert, Andreas and van Rossum, Mark CW}]{sacramento2015energy}
\textbf{\color{eLifeMediumGrey} Sacramento J}, Wichert A, van Rossum MC.
\newblock Energy efficient sparse connectivity from imbalanced synaptic plasticity rules.
\newblock PLoS Computational Biology.  2015; 11(6):e1004265.

\bibitem[{Sakaba and Neher(2001)Sakaba, Takeshi and Neher, Erwin}]{sakaba2001quantitative}
\textbf{\color{eLifeMediumGrey} Sakaba T}, Neher E.
\newblock Quantitative relationship between transmitter release and calcium current at the calyx of held synapse.
\newblock Journal of Neuroscience.  2001; 21(2):462--476.

\bibitem[{Schug et~al.(2021)Schug, Simon and Benzing, Frederik and Steger, Angelika}]{schug2021presynaptic}
\textbf{\color{eLifeMediumGrey} Schug S}, Benzing F, Steger A.
\newblock Presynaptic stochasticity improves energy efficiency and helps alleviate the stability-plasticity dilemma.
\newblock eLife.  2021; 10:e69884.

\bibitem[{Shannon(1948)Shannon, Claude Elwood}]{shannon1948mathematical}
\textbf{\color{eLifeMediumGrey} Shannon CE}.
\newblock A mathematical theory of communication.
\newblock The Bell system technical journal.  1948; 27(3):379--423.

\bibitem[{Shivkumar et~al.(2018)Shivkumar, Sabyasachi and Lange, Richard and Chattoraj, Ankani and Haefner, Ralf}]{shivkumar2018probabilistic}
\textbf{\color{eLifeMediumGrey} Shivkumar S}, Lange R, Chattoraj A, Haefner R.
\newblock A probabilistic population code based on neural samples.
\newblock Advances in neural information processing systems.  2018; 31.

\bibitem[{Silver(2003)Silver, R Angus}]{silver2003estimation}
\textbf{\color{eLifeMediumGrey} Silver RA}.
\newblock Estimation of nonuniform quantal parameters with multiple-probability fluctuation analysis: theory, application and limitations.
\newblock Journal of neuroscience methods.  2003; 130(2):127--141.

\bibitem[{Sj{\"o}str{\"o}m et~al.(2001)Sj{\"o}str{\"o}m, Per Jesper and Turrigiano, Gina G and Nelson, Sacha B}]{sjostrom2001rate}
\textbf{\color{eLifeMediumGrey} Sj{\"o}str{\"o}m PJ}, Turrigiano GG, Nelson SB.
\newblock Rate, timing, and cooperativity jointly determine cortical synaptic plasticity.
\newblock Neuron.  2001; 32(6):1149--1164.

\bibitem[{Turrigiano et~al.(1998)Turrigiano, Gina G and Leslie, Kenneth R and Desai, Niraj S and Rutherford, Lana C and Nelson, Sacha B}]{turrigiano1998activity}
\textbf{\color{eLifeMediumGrey} Turrigiano GG}, Leslie KR, Desai NS, Rutherford LC, Nelson SB.
\newblock Activity-dependent scaling of quantal amplitude in neocortical neurons.
\newblock Nature.  1998; 391(6670):892--896.

\bibitem[{Turrigiano and Nelson(2004)Turrigiano, Gina G and Nelson, Sacha B}]{turrigiano2004homeostatic}
\textbf{\color{eLifeMediumGrey} Turrigiano GG}, Nelson SB.
\newblock Homeostatic plasticity in the developing nervous system.
\newblock Nature reviews neuroscience.  2004; 5(2):97--107.

\bibitem[{Whittington and Bogacz(2017)Whittington, James CR and Bogacz, Rafal}]{whittington2017approximation}
\textbf{\color{eLifeMediumGrey} Whittington JC}, Bogacz R.
\newblock An approximation of the error backpropagation algorithm in a predictive coding network with local hebbian synaptic plasticity.
\newblock Neural computation.  2017; 29(5):1229--1262.

\bibitem[{Yang et~al.(2009)Yang, Guang and Pan, Feng and Gan, Wen-Biao}]{yang2009stably}
\textbf{\color{eLifeMediumGrey} Yang G}, Pan F, Gan WB.
\newblock Stably maintained dendritic spines are associated with lifelong memories.
\newblock Nature.  2009; 462(7275):920--924.

\bibitem[{Yang and Li(2022)Yang, Yukun and Li, Peng}]{yang2022synaptic}
\textbf{\color{eLifeMediumGrey} Yang Y}, Li P.
\newblock Synaptic Dynamics Realize First-order Adaptive Learning and Weight Symmetry.
\newblock arXiv preprint arXiv:221209440.  2022; .

\bibitem[{Yu et~al.(2016)Yu, Lianchun and Zhang, Chi and Liu, Liwei and Yu, Yuguo}]{yu2016energy}
\textbf{\color{eLifeMediumGrey} Yu L}, Zhang C, Liu L, Yu Y.
\newblock Energy-efficient population coding constrains network size of a neuronal array system.
\newblock Scientific reports.  2016; 6(1):19369.

\end{thebibliography}


\appendix
\begin{appendixbox}
\label{first:app}
\section{Reliability costs}
\hfill \break
The difficulty in determining reliability costs is that $\sigma$ depends on three variables: $n$, the number of vesicles, $p$ the probability of release and $q$, the quantal size, which measures the amount of neurotransmitter in each vesicle:
\begin{align}
\sigma^2 &= n p (1-p) q^2.
\end{align}
 However, these variables also determine the mean
\begin{equation}
\mu = n p q
\end{equation}  
so a straighforward optimisation of $\sigma$ under reliability costs will also change $\mu$ and one pitfall is to accidentally consider only those changes in reliability that derive from changes in mean. A solution to this problem is to eliminate one of the variables so that $\sigma$ is a function of $\mu$, and the remaining variables. Eliminating $q$ gives
\begin{equation}
\sigma = \mu \sqrt{\frac{1-p}{np}} 
\end{equation}
The idea is to consider energetic costs associated with $p$ and $n$ and relate these to $\sigma$ while holding $\mu$ fixed. To simplify the biological motivation, we assume that during changes to the synapse aimed at manipulating the energetic cost, $q$ will also change to compensate for any collateral changes in $\mu$ keeping $\mu$ constant. Hence, $q$ a ``compensatory variable''. Moreover, there is biological evidence that $q$ is the mechanism used in real synapses to fix $\mu$ \citep{turrigiano1998activity, karunanithi2002quantal}.
\add{Fixing $\mu$ through a compensatory variable is termed homeostatic plasticity. Fixing $\mu$ through $q$ is described as ``quantal scaling''. For reviews on homeostatic plasticity see \citet{turrigiano2004homeostatic, davis2015homeostatic}.}

In what follows four different energy costs are considered, the first depends on $p$, the next two on $n$, the situation for the final one is less clear, but in each case we derive a reliability cost in the form $\sigma^{-\rho}$ for some value of $\rho$. Of course, since we are considering costs for fixed $\mu$ the coefficient of variation $k=\sigma/\mu$ is proportional to $\sigma$; it may be helpful to think of these calculations as finding the relationship between the cost and $k$.

\subsection{Calcium efflux -- $\rho = \frac{1}{2}$}
 Calcium influx into the synapse is an essential part of the mechanism for vesicle release. Presynaptic calcium pumps act to restore calcium concentrations in the synapse; this pumping is a significant portion of synaptic transmission costs \citep{attwell2001energy}. By rearranging the Hill equation defined by \citealt{sakaba2001quantitative}, it can be shown that vesicle release has an interaction coefficient of four, this means the \emph{odds} of release per vesicle are related to intracellular calcium amplitude via the fourth power \citep{heidelberger1994calcium, sakaba2001quantitative}:
 \begin{equation}
     \frac{p}{1-p} \propto [Ca]^4.
 \end{equation}
To recover basal synaptic calcium concentration, the calcium influx is reversed by ATP-driven calcium pumps, where $[Ca]\propto ATP$: 
\begin{equation}
\text{Calcium pump cost} \propto \sqrt[4]{\frac{p}{1-p}}.
\end{equation}
Since this physiological cost does not depend on $n$ we assume it is fixed and so the odds of release $p/(1-p)$ is proportional to $\sigma^{-2}$. Thus
\begin{equation}
\text{Calcium pump cost} \propto \frac{1}{\sigma^{\frac{1}{2}}}
\end{equation}
or $\rho=1/2$.


\subsection{Vesicle membrane -- $\rho = \frac{2}{3}$}
There is a cost associated with the total area of vesicle membrane. Evidence in \citealt{pulido2021synaptic} suggest stored vesicles emit charged H\textsuperscript{+} ions, with the number emitted proportional to the number of v-glut, glutamate transporters, on the surface of vesicles. v-ATPase pumps reverse this process maintaining the pH of the cell. It is suggested that this cost is 44\% of the resting synaptic energy consumption. In addition,
metabolism of the phospholipid bilayer that form the membrane of neurotransmitter filled vesicles has been identified as a major energetic cost \citep{purdon2002energy}. Provided the total volume is the same, release of the same amount of neurotransmitter into the synaptic cleft can involve many smaller vesicles or fewer larger ones. However, while having many small vesicles will be more reliable, it requires a greater surface area of costly membrane. With fixed $\mu$ and $p$,
\begin{equation}
   \text{Vesicle membrane cost} \propto nr^2  
\end{equation}
Since $r^2\propto q^{2/3}$ and using $\mu=npq$ this give
\begin{equation}
    \text{Vesicle membrane cost} \propto \left(\frac{n\mu^2}{p^2}\right)^{1/3}
\end{equation}
Since this reliability cost depends on $n$, so $p$ is regarded as constant, so $\sigma^{-2}\propto n$ and hence $\rho=2/3$.

\subsection{Actin -- $\rho = \frac{4}{3}$}
Actin polymers are an energy costly structural filament that support the structural organisation of vesicle reserve pools \citep{cingolani2008actin}, with the vesicles strung out along the filamants. We assume each vesicle to require a length of actin roughly proportional its diameter, this means that the total length of actin is proportional to $nr$. Hence
\begin{align}
\text{Actin cost} \propto nr
\end{align}
The calculation then proceeds much as for the membrane cost, but with $r$ instead of $r^2$ giving $\rho=4/3$.

\subsection{Trafficking -- $\rho = 2$}
ATP fueled myosin motors drive trains of actin filament along with associated cargo such as vesicles and actin-myson trafficking moves vesicles from vesicle reserve pools to release sites sustaining the readily releasable pool (RRP) following vesicle release \citep{bridgman1999myosin, gramlich2017actin}. This gives a cost for vesicle recruitment proportional to $np$, the number of vesicles released:
\begin{align}
\text{Trafficking cost} \propto np
\end{align}
so if $n$ is regarded as the principal way this cost is changed, with $p$ fixed then $n\propto \sigma ^{-2}$ and so $\rho=2$. This is the view point we are taking to motivate examining reliability costs with $\rho=2$. This is certainly useful in considering the range of model behaviours over a broad a range of $\rho$ values. 

Nonetheless, it is sensible to ask whether the likely biological mechanism behind a varying trafficking cost is one which changes $np$ itself. In this case, since a constant $\mu$ for varying $np$ means $q\propto 1/np$ we have
\begin{align}
\text{Trafficking cost} \propto \frac{1-p}{\sigma^2}
\end{align}
which, is, again, of the form cost $\propto \sigma^{-2}$ provided $p$ is small. For larger $p$, however, it depends on exactly how $p$ changes as $np$ changes.

Generally, throughout these calculation we have supposed that $q$ is a compensatory variable and that either $p$ or $n$ changes in the process that changes the cost at a synapse. The benefit of this is that we are able to model costs directly in terms of reliability; in the future, though, it might be interesting to consider models which use $n$, $p$ and $q$ instead of $\mu$ and $\sigma$; this would certainly be convenient for the sort of comparisons we are making here and interesting, although two formulations seem equivalent, this does not mean that the learning dynamics will be the same.
\end{appendixbox}

\begin{appendixbox}
\section{High input rates and high precision at important synapses}
\label{second:app}
\hfill \break
Here, we show that under a linear model, important synapses, as measured by the Hessian, have high input rates.
Additionally, we show that energy efficient synapses implies that these important synapses have low optimized variability.

We consider a simplified linear model, with targets $\bm{y}$, inputs $\bm{X}$ and weights $\bm{w}$.
We have,
\begin{equation}
  \P(\bm{y}| \bm{w}, \bm{X}) = \mathcal{N}(\bm{y}; \bm{X} \bm{w}, \epsilon^2 \bm{I}).
\end{equation}
We take the performance cost to be,
\begin{align}
\label{eq:performance_cost_P}
\text{performance cost} = -\log \P(\bm{y}| \bm{w}, \bm{X}).
\end{align}
where $N$ is the number of datapoints.
We take the network weights to be drawn from a multivariate Gaussian, with diagonal covariance, $\bm{\Sigma}$, i.e.\ $\Sigma_{ii} = \sigma_i^2$,
\begin{equation}
  \Q(\bm{w}) = \mathcal{N}(\bm{w}; \bm{\mu},\bm{\Sigma}).
\end{equation}
All our derivations rely on looking at the quadratic form for the performance cost.
In particular,
\begin{equation}
\text{performance cost} = \E_{\Q(\bm{w})}\left[\frac{1}{2\epsilon^2} (\bm{y}-\bm{X w})^T(\bm{y}-\bm{X w})\right]
\end{equation}
Expanding the brackets,
\begin{equation}
\text{performance cost}=
\frac{1}{2\epsilon^2}(\bm{y}^T \bm{y} - 2 \bm{y}^T \bm{X}\E[w] + E[\bm{w}^T \bm{X}^T \bm{X w}]),
\end{equation}
and evaluating the expectations,
\begin{equation}
\label{eq:perf_cost}
\text{performance cost}=
\frac{1}{2\epsilon^2}[\bm{y} \bm{y}^T - 2 \bm{y}^T \bm{X}\bm{\mu} + \tr(\bm{X}^T \bm{X} \bm{\Sigma}) + \bm{\mu}^T\bm{X}^T \bm{X} \bm{\mu }]
\end{equation}
where the trace,
\begin{equation}
    \tr(\bm{X}^T \bm{X} \bm{\Sigma}) = \sum_i\sum_t X_{ti}^2\sigma_i^2,
\end{equation}
includes a sum over both the synapse index $i$ and the time index $t$.

To measure synapse importance, we use the Hessian, that is the second derivative of the performance cost with respect to the mean. 
From Eq.~\eqref{eq:perf_cost}, we can identify this as,
\begin{align}
  \label{eq:H}
  H_{ii} &= \frac{\partial^2 \text{performance cost}}{\partial \mu_i^2} = \frac{1}{\epsilon^2} \sum_t X_{ti}^2.
\end{align}
where, as before, $t$ is the time index.

Since the the only term in the performance cost that depends on the variance, $\sigma^2_i$, is the trace, we can identify $\sigma_i$ that optimises the performance and reliability cost. This allows us to observe how synapse variability relates to synapse importance in the tradeoff between performance and reliability costs:
\begin{equation}
    \frac{d}{d\sigma_i}[\text{performance cost} + \text{reliability cost}] = H_{ii} \sigma_i -s^{\rho}\sigma_i^{-\rho-1}=0  
\end{equation}
which means
\begin{equation}
\sigma_i^{\rho+2} = s^{\rho}/H_{ii}
\end{equation}
or
\begin{equation}
\label{eq:sigma_energy}
\sigma_i =  \sqrt[\rho+2]{s^\rho/H_{ii}} 
\end{equation}
Thus, more important synapses (as measured by the Hessian, $H_{ii}$) have lower variability if the synapse is energy efficient.
Moreover, through Eq.~\eqref{eq:H}, we expect synapses with higher input rates to be more important.
\end{appendixbox}

\begin{appendixbox}
\section{Synapse importance and gradient magnitudes}
\label{third:app}
\hfill \break
In our ANN simulations we train synaptic weights using the most established adaptive optimisation scheme, Adam \citep{kingma2014adam}; which has recently been realised using biologically plausible mechanisms \citep{yang2022synaptic}.
Adam uses a synapse-specific learning rate, $\eta_i$, which decreases in response to high gradients at that synapse,
\begin{align}
  \eta_i &= \frac{\eta_\text{base}}{\sqrt{\langle g_i^2 \rangle}}.
  \label{eq:lr}
\end{align}
Specifically, the local learning rate for the $i$th synapse, $\eta_i$, is usually a base learning rate, $\eta_\text{base}$, divided by $\sqrt{\langle g_i^2 \rangle}$, the root-mean-square of the gradient for each datapoint or minibatch.

The key intuition is that if the gradients for each datapoint/minibatch are large, that means that this synapse is important, as it has a big impact on the predictions for every datapoint.
In fact, these mean-square gradients can be related to our formal measure of synapse importance, the Hessian.
Specifically, for data generated from the model, the Hessian (or Fisher information \citealp{fisher1922mathematical}) is equivalent to the mean squared gradient, where the gradient is taken over each datapoint separately,
\begin{align}
  \label{eq:fi}
  E_{P(y, \bm{x}|\bm{w})}\left[\frac{\partial^2 \text{performance cost}}{\partial \bm{\mu}^2}\right] = -E_{P(y, \bm{x}|\bm{w})}\left[\frac{\partial^2 \log P(y| \bm{w}, \bm{x})}{\partial \bm{\mu}^2}\right] = E_{P(y, \bm{x}|\bm{w})}[\bm{g}\bm{g}^T],
\end{align}
\add{where the expectation is evaluated over data generated by the model and $\bm{g}$ is defined,}
\begin{equation}
\bm{g} = \frac{\partial}{\partial \bm{\mu}}\log P(y| \bm{w}, \bm{x})
\end{equation}
Of course, in practice, the data is not drawn from the model, so the relationship between the squared gradients and the Hessian computed for real data is only approximate.
But it is close enough to induce a relationship between synapse importance (measured as the diagonal of the Hessian) and learning rates (which are inversely proportional to the root-mean-square gradients) in our simulations.

\end{appendixbox}

\begin{appendixbox}
\section{Energy efficient noise and variational-Bayes for neural network weights}
\label{fourth:app}
\hfill \break
\textbf{Introduction to Variational Bayes for neural network weights} One approach to performing Bayesian inference for the weights of a neural network is to use variational Bayes.
In variational Bayes, we introduce a parametric approximate posterior, $Q(w)$, and fit the parameters of that approximate posterior using gradient descent on an objective, the ELBO \citep{blundell2015weight}.
In particular, 
\begin{align}
  \label{eq:elbo}
  \log P(y| x) \geq \text{ELBO} &= E_{Q(w)}[\log P(y| x, w) + \log P(w)] + H[Q(w)].
\end{align}
where $H[Q(w)]$ is the entropy of the approximate posterior.
Maximising the ELBO is particularly useful for selecting models as it forms a lower-bound on the marginal-likelihood, or evidence, $\log P(y| x)$ \citep{mackay1992evidence, barber1998ensemble}.
When using variational Bayes for neural network weights, we usually use Gaussian approximate posteriors \citep{blundell2015weight},
\begin{align}
  Q(w_i) &= \mathcal{N}(w_i; \mu_i, \sigma_i^2).
\end{align}
Note that optimising the ELBO wrt the parameters of $Q$ is difficult, because $Q$ is the distribution over which the expectation is taken in Eq.~\eqref{eq:elbo}.
To circumvent this issue, we use the reparameterisation trick \citep{kingma2015variational}, which involves writing the weights in terms of IID standard Gaussian variables, $\epsilon$,
\begin{align}
  w_i &= \mu_i+ \sigma_i \epsilon_i.
\end{align}
Thus, we can write the ELBO as an expectation over $\epsilon$, which has a fixed IID standard Gaussian distribution,
\begin{align}
  \text{ELBO} &= E_{\epsilon}\left[\log P(y| x, w{=}\mu{+}\sigma \epsilon) + \log P(w{=}\mu{+}\sigma \epsilon)\right] + H[Q(w)].
  \label{eq:elbo2}
\end{align}
Note that we use $w$, $\mu$ etc. without indicies in this expression to indicate all the weights / mean weights.

\textbf{Identifying the log-likelihood and log-prior.}
Following the usual practice in deep learning, we assume the likelihood is formed by a categorical distribution, with probabilities obtained by applying the softmax function to the output of the neural network.
The log-likelihood for a categorical distribution with softmax probabilities is the negative cross-entropy \citep{murphy2012machine},
\begin{align}
  \log P(y| x, w) &= -\text{cross entropy}(y; f(x, w))
\end{align}
where $f(x, w)$ is the output of the network with weights, $w$ and inputs, $x$.
We can additionally identify a Laplace prior with the magnitude cost.
Specifically, if we take the prior to be Laplace, with scale $1/\lambda$,
\begin{align}
  \log P(w) &= \sum_i \log \text{Laplace}(w_i; 0, \tfrac{1}{\lambda})\\
  &= -N \log (2/\lambda) - \lambda \sum_i |w_i|\\
  &= \text{const} - \text{magnitude cost}
\end{align}
where we identify the magnitude cost using Eq.~\eqref{eq:magnitudeCost}.

\textbf{Connecting the entropy and the biological reliability cost.}
Now, we have identified the log-likelihood and log-prior terms in the ELBO (Eq.~\ref{eq:elbo}) with terms in our biological cost (Eq.~\ref{eq:wordobjective}).
We thus have two terms left: the entropy in the ELBO (Eq.~\ref{eq:elbo}) and the reliability cost in the biological cost (Eq.~\ref{eq:wordobjective}).
Critically, the entropy term also acts as a reliability cost, in that it encourages more variability in the weights (as the entropy is positive in the ELBO (Eq.~\ref{eq:elbo}) and we are trying to maximise the ELBO, the entropy term gives a bonus for more variability).
Intuitively, we can think of the entropy term in the ELBO (Eq.~\ref{eq:elbo}) as being an ``entropic reliabity cost'', as compared to the ``biological reliability cost'' in (Eq.~\ref{eq:wordobjective}).

This intuitive connection suggests that we might be able to find a more formal link.
Specifically, we can define an entropic reliability cost as simply the negative entropy,
\begin{align}
 \Cvi &= -H[Q_i] = -\tfrac{1}{2} \ln 2 \pi e \sigma_i^2.
  \intertext{Our goal is to write the biological reliability cost, $\Cbi$, as a bound on $\Cvi$. By rearranging and introducing $s$ and $\rho$,}
  \Cvi &=  \frac{1}{\rho} \ln \b{\frac{s}{\sigma_i}}^{\rho} - \tfrac{1}{2} \ln 2\pi e s^2, 
  \end{align}
  and noting that $\log a \leq a-1$,
  \begin{equation}
  \log \b{\frac{s}{\sigma}}^{\rho} \leq \b{\frac{s}{\sigma_i}}^{\rho} -1
  \end{equation}
 we can demonstrate that any reliability cost expressed as a generic power-law forms an upper bound on the entropic cost,
  \begin{align}
  \label{eq:Cbi}
 \Cvi &\leq \frac{1}{\rho} \b{\b{\frac{s}{\sigma_i}}^{\rho} -1} - \tfrac{1}{2} \log 2\pi e s^2 = \Cbi = \text{reliability cost}_i + \text{const}_i.
\end{align}
\add{Here, $\Cbi$ is the biological reliability cost for a synapse, which formally bounds the entropic reliability cost from VI,}
\begin{align}
  \text{const}_i &= -\tfrac{1}{\rho} - \tfrac{1}{2} \log 2\pi e s^2 & 
  \text{reliability cost}_i &= \frac{1}{\rho} \b{\frac{s}{\sigma_i}}^{\rho} = c \sigma_i^{-\rho}
\end{align}
\add{We can also sum these quantities across synapses,}
\begin{align}
  \text{const} &= \sum_i \text{const}_i &
  \text{reliability cost} &= \sum_i \text{reliability cost}_i.\\
 \Cv &= \sum_i\Cvi & \Cb &= \sum_i \Cbi.
\end{align}
The parameter $c$ sets the importance of the reliability cost within the performance-reliability cost tradeoff (see Fig.~\ref{fig:accuracy}) 
\begin{align}
\label{eq:csrho}
c = s^\rho/\rho.
\end{align}
Note that while both $\text{reliability cost}_i$ and $\Cbi$ represent the biological reliability costs, they are slightly different in that $\Cbi$ includes an additive constant. 
Importantly, this additive constant is independent of $\sigma_i$ and $\mu_i$, so it does not affect learning and can be ignored. 

Given that the biological reliability cost forms a bound on the ideal entropic reliability cost we can consider using the biological reliability cost in place of the entropic reliability cost,
\begin{align}
  \log P(y| x) \geq \text{ELBO} &= E_{Q(w)}[\log P(y| x, w) + \log P(w)] - \Cv \\
  \nonumber
  \log P(y| x) \geq \text{ELBO} &\geq E_{Q(w)}[\log P(y| x, w) + \log P(w)] - \Cb = -\text{overall cost} - \const
\end{align}
we find that our overall biological cost (Eq.~\ref{eq:wordobjective}) forms a bound on the ELBO, which itself forms a bound on the evidence, $\log P(y| x)$. 
Thus, pushing down the overall biological cost (Eq.~\ref{eq:wordobjective}) pushes up a bound on the model evidence.

\textbf{Predictive probabilities arising from biological reliability costs}
Given the connections between our overall cost and the ELBO, we expect that optimising our overall cost will give a similar result to variational Bayes. To check this connection, we plotted the distribution of predictions induced by noisy weights arising from variational Bayes (Appendix 4--Fig.~\ref{fig:evidence}a) and our overall costs (Appendix 4--Fig.~\ref{fig:evidence}b). 
Variational Bayes maximises the ELBO, therefore its predictive distribution is optimised to reflect the data distribution from which data is drawn \citep{mackay1992practical}. 
We found comparable patterns for the predictive distributions learned through variational Bayes and our overall costs, albeit with some breakdown in predictive performance with higher values for $\rho$.
\begin{figure}[H]
\includegraphics[width=1\textwidth]{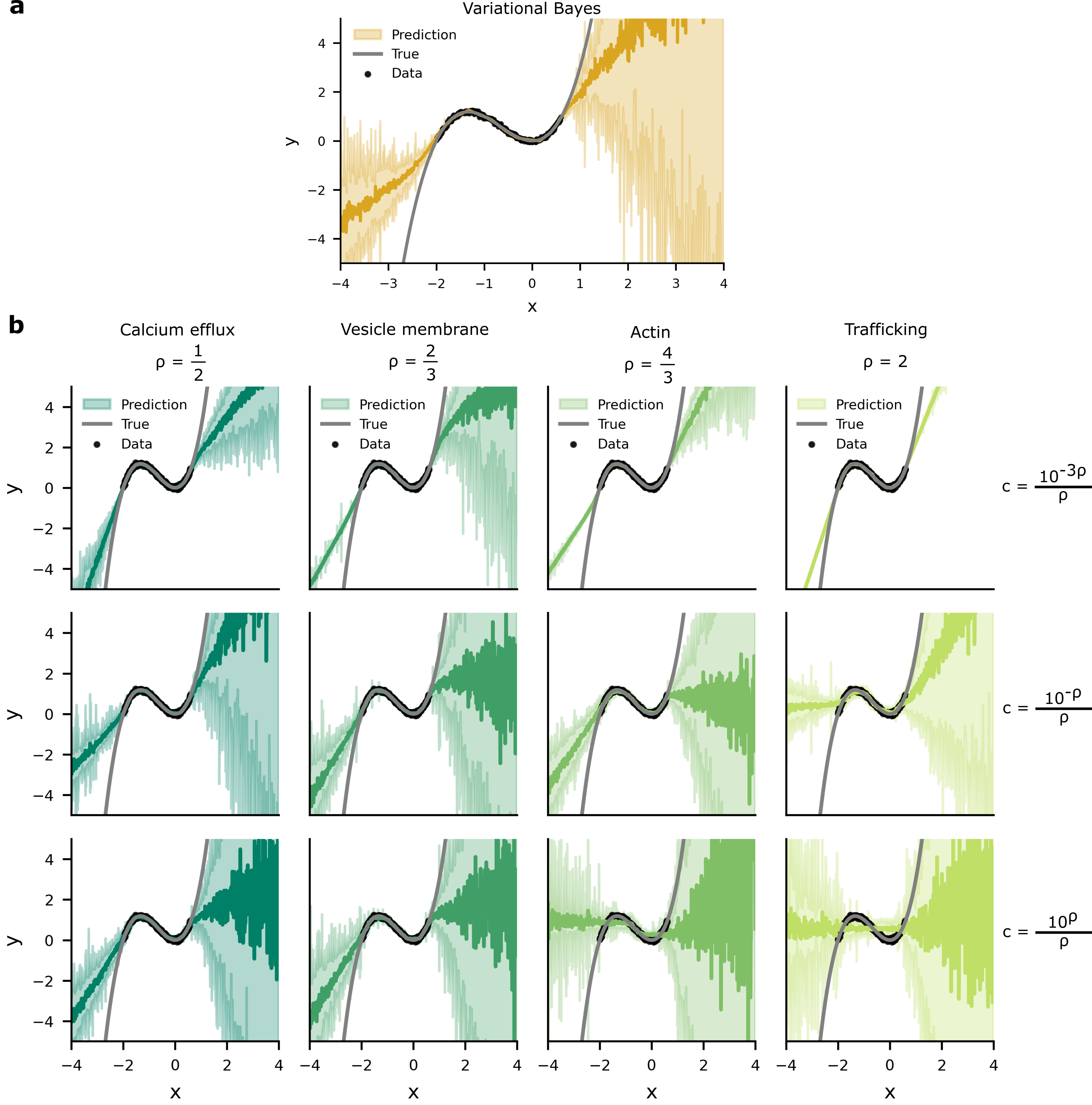}
\caption{\textbf{Predictive distributions from variational Bayes and our overall costs are similar}.  We trained a one hidden layer network with $20$ hidden units on data $(y_a,x_a)$ (black dots). Network targets, $y_a$, were drawn from a ``true'' function, $f(x)=x^3 + 2x^2$ (grey line) with additive Gaussian noise of variance $0.05^2$. 
Two standard deviations of the predictive distributions are depicted in the shaded areas. \textbf{a.} The predictive distribution produced by variational Bayes have a larger density of predictions where there is a higher probability of data. Where there is an absence of data, the model has to extrapolate and the spread of the predictive distribution increases. \textbf{b.} Optimising the overall cost with small $c$ generate narrow response distributions. This is most noticeable in the upper-right panel; where the spread of predictive distribution is unrelated to the presence or absence of data. In contrast, while the predictive density for larger $c$ do vary according to the presence or absence of data, these distributions poorly predict $f(x)$. This is most apparent in the lower-right panel where the network's predictive distribution transects the inflections of $f(x)$. 
\label{fig:evidence}
}
\end{figure}
\textbf{Interpreting $c$, $\rho$ and $s$}
As discussed in the main text, $c$ and $\rho$ are the fundamental parameters, and they are set by properties of the underlying biological system.
It may nonetheless be interesting to consider the effects of $s$ and $\rho$ on the tightness of the bound of the biological reliability cost on the variational reliability cost.
In particular, we consider settings of $s$ and $\rho$ for which the bound is looser or tighter (though again, there is no free choice in these parameters: they are set by properties of the biological system).

First, the biological reliability cost becomes equal to the ideal entropic reliability cost in the limit as $\rho \rightarrow 0$.
\begin{align}
  \lim_{x \rightarrow 0} \tfrac{1}{x} (z^x - 1) = \log z
\end{align}
Thus, taking $\rho = x$, and $z = s/\sigma_i$,
\begin{align}
  \lim_{\rho \rightarrow 0} \tfrac{1}{\rho} \b{\b{\tfrac{s}{\sigma_i}}^{\rho} - 1} = \log \tfrac{s}{\sigma_i}
  \label{eq:smallrho}
\end{align}
Thus,
\begin{align}
  \lim_{\rho \rightarrow 0} \Cbi &= \log \tfrac{s}{\sigma_i} - \tfrac{1}{2} \log 2 \pi e s^2 = - \tfrac{1}{2} \log 2 \pi e \sigma_i^2 = \Cvi.
\end{align}
This explains the apparent improvement in predictive performance (Appendix 4--Fig.~\ref{fig:evidence}) and in matching the posteriors (Fig.~\ref{fig:geometric}) with lower values of $\rho$.

Second, the biological reliability cost becomes equal to the ideal entropic reliability cost when $s=\sigma_i$, 
\begin{align}
  \Cbi(s=\sigma_i) &= - \tfrac{1}{2} \log 2 \pi e s^2 = - \tfrac{1}{2} \log 2 \pi e \sigma_i^2 = \Cvi.
\end{align}
as the first term in Eq.~\ref{eq:Cbi} cancels.
However, $s$ cannot be set individually across synapses, but is instead roughly constant, with a value set by underlying biological constraints. 
In particular, $s$ can be written as a function of $\rho$ and $c$ (Eq.~\ref{eq:csrho}), and $\rho$ and $c$ are quantities that are roughly constant across synapses, with their values set by biological constraints.
Thus, biological implications of a tightening bound as $s$ tends to $\sigma_i$ are not clear.
\end{appendixbox}

\begin{appendixbox}
\section{Analytic predictions for $\sigma_i$}
\label{fifth:app}
\hfill \break
At various points in the main text, we note a connection between the Hessian, synapse importance and optimal variability.
We start with Eq.~\eqref{eq:sigma_energy}, which relates the optimal, energy efficient noise variance, $\sigma_i^2$ to the Hessian, $H_{ii}$ which gives Eq.~\eqref{eq:sigma_eta}.
Then, we combine this with the form for the Hessian (Eq.~\ref{eq:H}), which gives Eq.~\eqref{eq:sigma_x}.
Finally, we note Hessian describes the log-likelihood (Eq.~\ref{eq:H} and Eq.~\ref{eq:performance_cost_P}).  Thus, assuming the prior variance is large, we have $H_{ii} = \sigma_{\text{post};i}$, which gives Eq.~\eqref{eq:sigma_sigma},
\begin{subequations}
\label{eq:identities}
\begin{align}
  \label{eq:sigma_eta}
  H_{ii} &\propto \langle g_i^2 \rangle \propto \tfrac{1}{\eta_i^2} & \log H_{ii} &= -2 \log \eta_i + \const & \log \sigma_i^2 &= \tfrac{4}{\rho + 2} \log \eta_i + \const\\
  \label{eq:sigma_x}
  H_{ii} &\propto \langle x_{i}^2 \rangle & \log H_{ii} &= \log \langle x_{i}^2 \rangle &  \log \sigma_i^2 &= -\tfrac{2}{\rho + 2} \log \langle x_{i}^2 \rangle  + \const\\
  \label{eq:sigma_sigma}
  H_{ii} &\propto \tfrac{1}{\sigma_{\text{post}; i}^2} & \log H_{ii} &= - \log \sigma_{\text{post}; i}^2 +\const & \log \sigma_i^2 &= \tfrac{2}{\rho+2} \log \sigma_{\text{post}; i}^2 + \const.
\end{align}
\end{subequations}
To test these predictions, we performed a simpler simulation classifying MNIST in a network with no hidden layers.
We found that the analytic results closely matched the simulations, and that the biological slopes tend to match the $\rho=2$ better than the other values for $\rho$.
However, while the direction of the slope was consistent in deeper networks, the exact value of the slope was not consistent (Fig.~\ref{fig:optimising} and Supplementary - Appendix 6--Fig.~\ref{fig:all_layers}), so it is unclear whether we can draw any strong conclusions here.

\begin{figure}[H]
\includegraphics[width=1\textwidth]{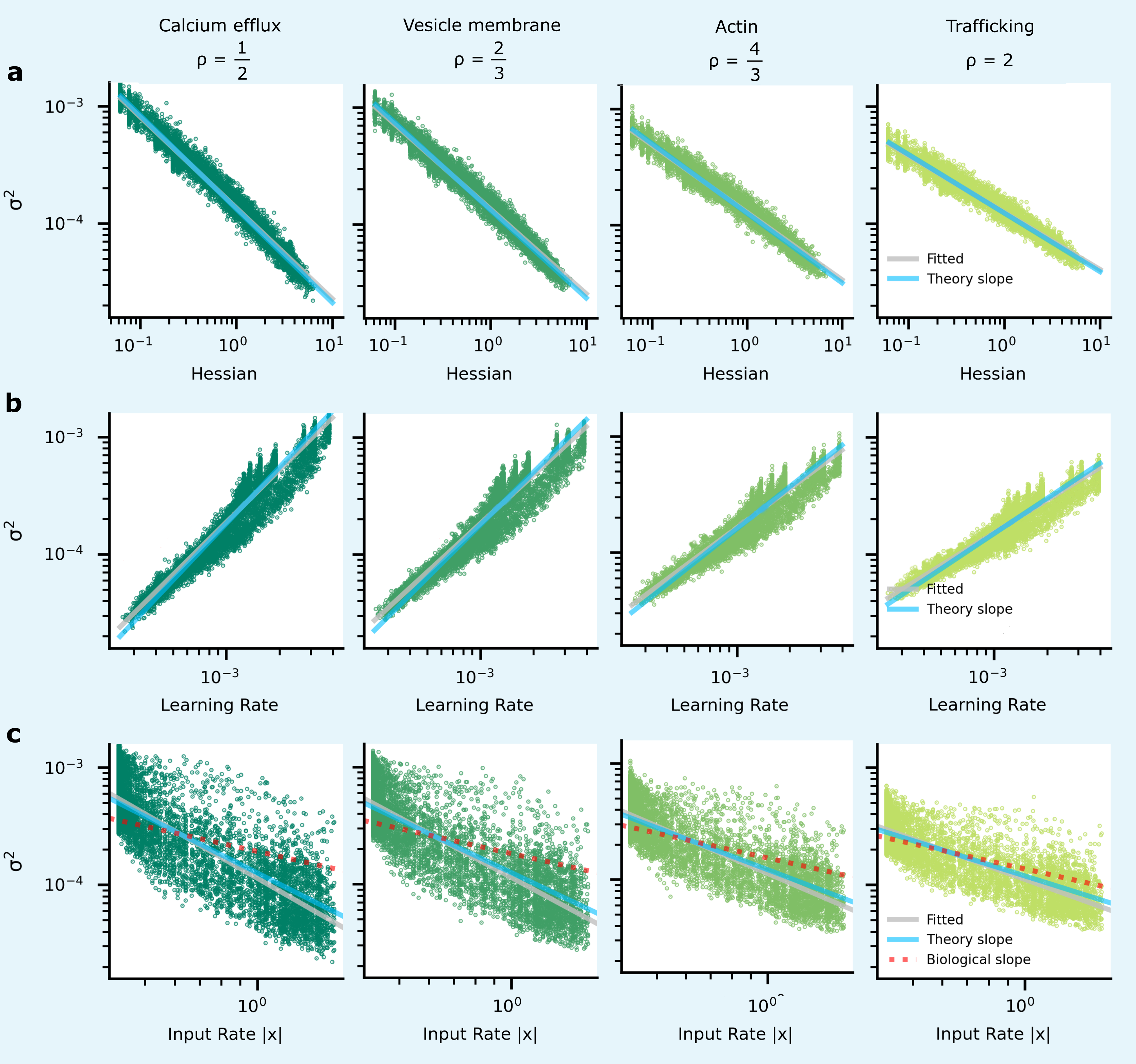}
\caption{\textbf{Comparing analytic predictions for synapse variability with simulations and experimental data in a zero-hidden-layer network for MNIST classification.} The green dots show simulated synapses, and the grey line is fitted to these simulated points. The blue line is from our analytic predictions, while the red dashed line is taken from experimental data (Fig.~\ref{fig:exp}b).
\label{fig:slopes}}
\end{figure}

\end{appendixbox}

\begin{appendixbox}
\section{Supplementary figures}
\label{sixth:app}
\begin{figure}[H]
\includegraphics[width=1\textwidth]{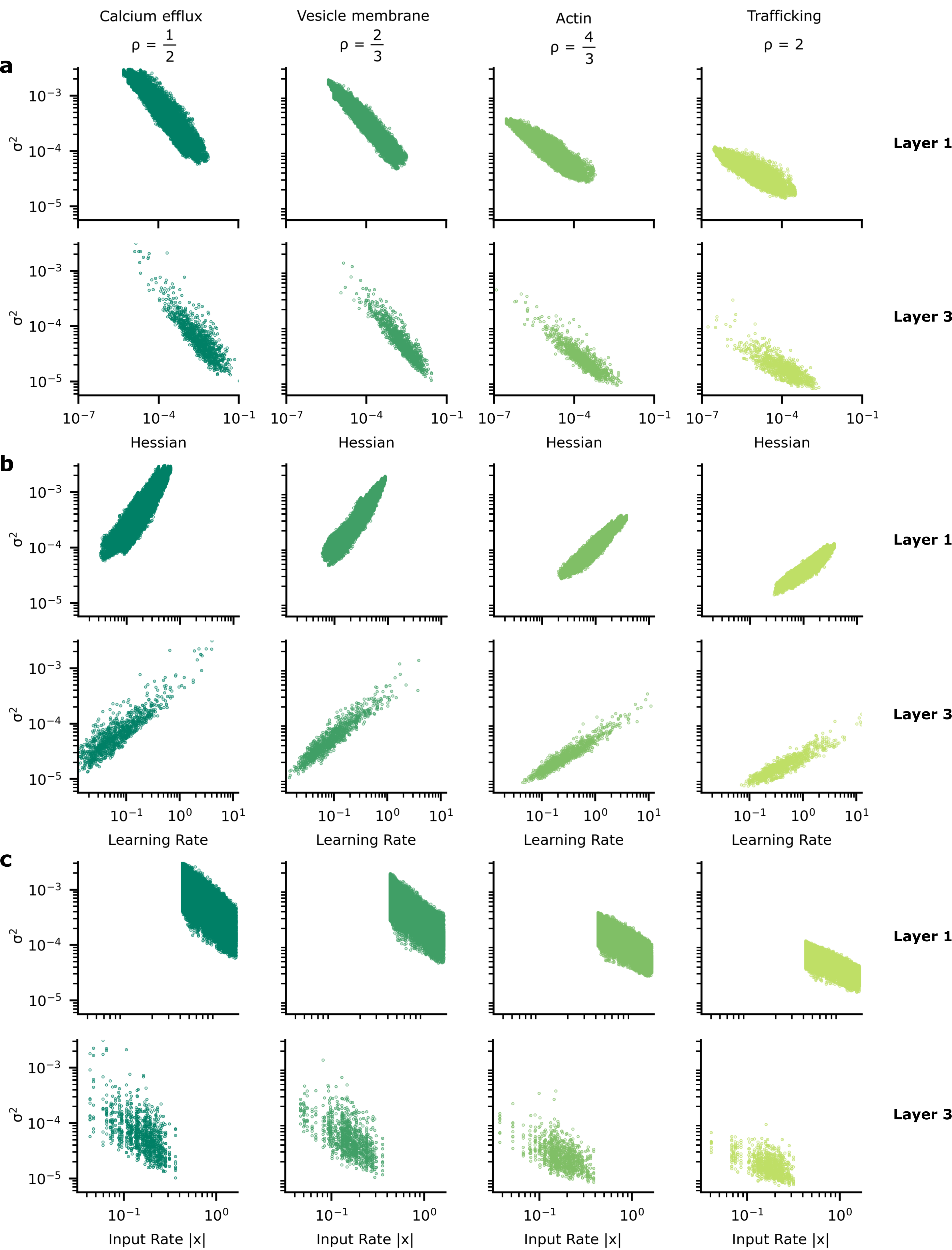}
\caption{\textbf{Patterns of synapse variability for the remaining layers of the neural network used to provide our results.} In Fig.~\ref{fig:optimising} we showed the heterogeneous patterns of synapse variability for the synapses connecting the two hidden layers of our ANN. Here we exhibit the equivalent plots for the other synapses, those between the input and the first hidden layer (layer 1) and \remove{the} from the final hidden layer to the output layer (layer 3). As in Fig.~\ref{fig:optimising} we show the relationship between synaptic variance and \textbf{a.} the Hessian; \textbf{b.} learning rate; and \textbf{c.} input rate. The patterns do not appear substantially different from layer to layer. \label{fig:all_layers}}
\end{figure}

\begin{figure}[H]  

\includegraphics[width=0.86\textwidth]{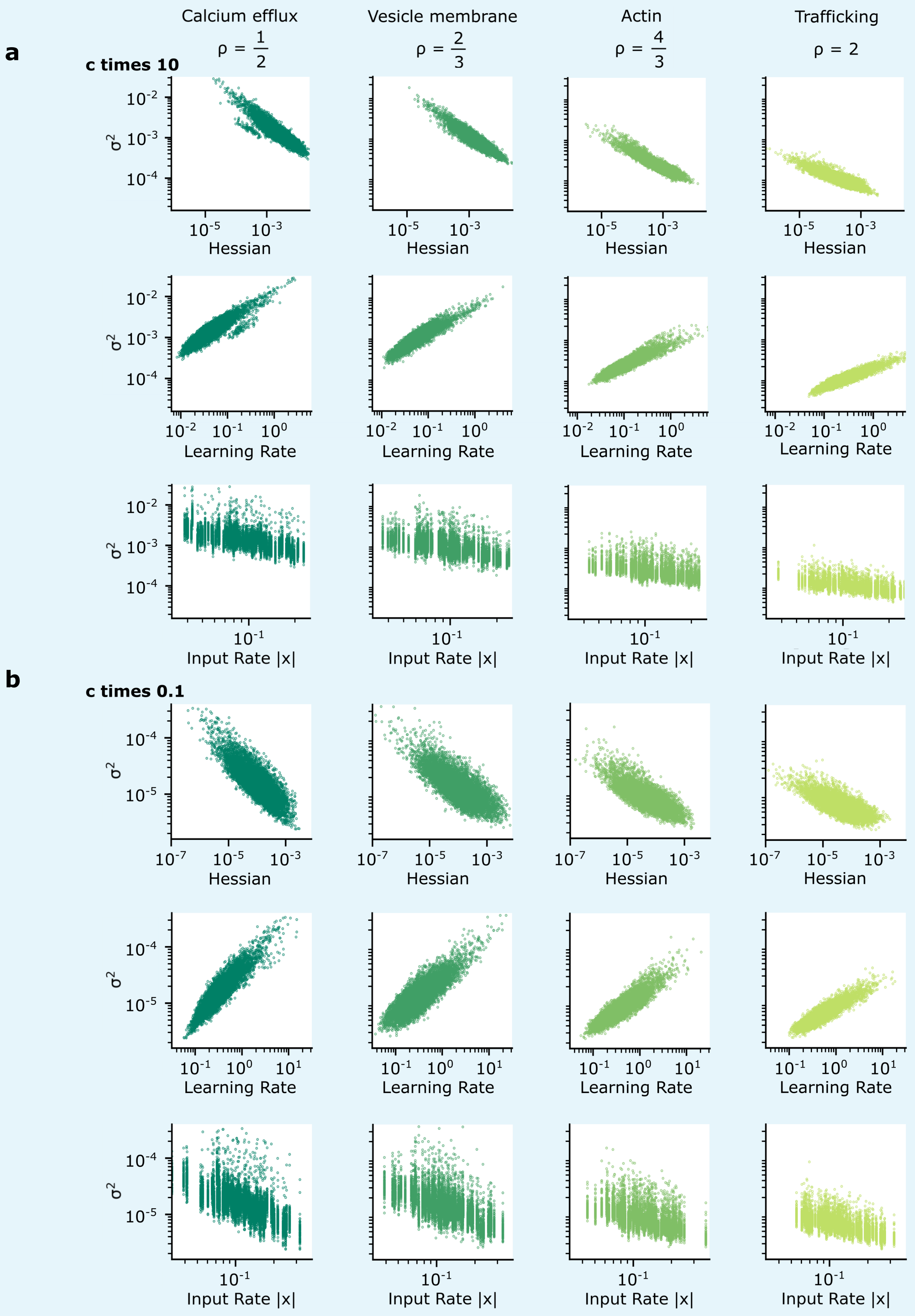}
\caption{\textbf{Patterns of synapse variability are robust to changes in the reliability.} We show that the patterns of variability of synapses connecting the two hidden layers presented in Fig.~\ref{fig:optimising} are preserved over a wide range of $c$. \textbf{a,b} When the reliability cost multiplier, $c$, is either increased (\textbf{a}) or decreased (\textbf{b}) by a factor of ten, overall synapse variability increases or decreases accordingly, but the qualitative correlations seen in Fig.~\ref{fig:optimising} are preserved.   
\label{fig:varying_c}}
\end{figure}
\end{appendixbox}

\end{document}